\documentclass[envcountsame]{llncs}

\usepackage{cite}
\usepackage{amsmath}
\usepackage{amssymb}
\usepackage{graphicx}
\usepackage{textcomp}
\usepackage{xcolor}
\usepackage{xspace}

\usepackage{listings}
\usepackage{hyperref}
\usepackage[export]{adjustbox}
\usepackage[caption=false]{subfig}
\usepackage{multirow}
\usepackage[linesnumbered,ruled,vlined]{algorithm2e}
\usepackage{mathtools}
\usepackage{wrapfig}
\usepackage{booktabs}
\usepackage{here}
\usepackage{mdframed}
\usepackage{enumitem}
\usepackage{dirtree}

\usepackage{article}

\lstdefinestyle{shell}{%
  basicstyle=\ttfamily\small,
  columns=flexible,
  keepspaces=true,
  showspaces=false,
  showstringspaces=false,
  showtabs=false,
  breaklines=true,
  breakautoindent=true,
  numbers=none,
  frame=single,
  rulecolor=\color{black!30},
  xleftmargin=8pt,
  xrightmargin=8pt,
  framexleftmargin=6pt,
  morecomment=[l]{....},
  commentstyle=\itshape\color{clComments},
  moredelim=[is][\color{clKeys}]{@}{@},
}

\begin{document}

\mainmatter

\title{Verifying Autoware Controller Node by Combining Model Verification and\\ Runtime Monitoring}

\author{Daisuke Ishii}
\institute{Japan Advanced Institute of Science and Technology}

\maketitle

\begin{abstract}
Formal modeling and verification of autonomous driving systems remain challenging due to the vast operational space and the complexity of the control logic.
We propose a hybrid verification approach that combines formal model verification and runtime monitoring for the trajectory-following control node of Autoware, an open-source autonomous driving platform.
We formalize the controller as a hybrid system in differential dynamic logic~(dL), and introduce an abstract model that captures the essential control dynamics.
We then prove safety and liveness properties of the model using the {KeYmaera~X} proof assistant.
Complementarily, we monitor the constraints that assure the compatibility between the model and the {Autoware} implementation across numerical simulation runs using the {Autoware} toolchain, quantifying to what extent the implementation's behavior is subsumed by the verified model.
The experimental results confirmed that (i)~monitoring provides a sound evaluation of target following with accuracy bounds, and (ii)~the Autoware implementation reaches the target with an accuracy of approximately $0.2$--$0.4$\,(m) in position and $0.8$\,(m/s) in velocity.
\end{abstract}

\section{Introduction}

Autonomous vehicles must operate safely and reliably in real-world environments, where a single failure can endanger lives and undermine public trust in the technology%
~\cite{othmanPublic2021}.
Among the core components of an \emph{autonomous driving system (ADS)}, the \emph{trajectory-following controller} (Sect.~\ref{s:tf})---responsible for computing steering and acceleration commands to track a planned trajectory---plays a central role in safe vehicle operation.

Rather than running vehicles over long distances to verify their safety~\cite{kalraDriving2016},
providing formal or quantitative assurances for ADS including controller components is of high practical importance, as recommended by ISO~26262 and ISO~21448.
However, due to the vast workspace and implementation complexity, most existing verification methods are currently limited to scenario- and simulation-based testing~\cite{tangCollision2021,viswanadhaAddressing2021,zhouSpecificationbased2023,tranSafety2025}.
Formal verification methods either rely on abstract models~\cite{rizaldiFormalising2017,tulcanHybrid2026} or runtime monitoring~\cite{schurmannEnsuring2017,nigamAutomating2023}, or target only a part of the ADS implementation~\cite{fritzschExperiences2021,konigSafe2024,nenchevCodeLevel2025}.
The combined complexity of the architecture and continuous vehicle dynamics makes formal modeling and verification of ADS implementations technically challenging.

\Impl{Autoware}~\cite{katoAutoware2018} (Sect.~\ref{s:autoware}) is an open-source ADS implementation widely used in research, prototyping, and real-world deployment.
It is based on ROS~2 and its toolchain includes a scenario-based simulation tool 
and game engine-based vehicle simulators.
Other than monitoring for ROS~\cite{huangROSRV2014,ferrandoROSMonitoring2020a} and scenario-based testing~\cite{tranSafety2025}, formal verification case studies for \Impl{Autoware} remain limited.

\emph{Hybrid verification methods} that combine formal model-based reasoning with runtime monitoring have been proposed as an effective approach to practical cyber-physical systems~\cite{luckcuckFormal2020,mehdipourFormal2023a}.
\emph{Differential dynamic logic~(dL)}~\cite{platzerLogical2018} and the \Impl{KeYmaera~X} proof assistant~\cite{fultonKeYmaera2015} provide a foundation for specifying and verifying hybrid systems with continuous dynamics; the \emph{ModelPlex} framework~\cite{mitschModelPlex2016} extends this to synthesize runtime monitors.
There are applications of ModelPlex to autonomous vehicles~\cite{bohrerFormal2019,linSafe2022,prebetVerification2025};
however, they were evaluated only with simple simulators and not with a practical ADS.
%
Formal models of the trajectory-following controller---involving the kinematic bicycle model, pure-pursuit lateral control, and PID longitudinal control---have not been developed, and how to monitor an ADS implementation like \Impl{Autoware} based on a dedicated formal model remains an open problem.
\Mod{How to adapt an existing model (e.g. \cite{bohrerFormal2019}) to such an implementation and conduct a hybrid verification is not straightforward.}

In this paper, we propose a hybrid verification approach (Sect.~\ref{s:verif}) for the trajectory-following controller node of \Impl{Autoware}, which can be regarded as a manual implementation of ModelPlex that combines formal model verification with runtime monitoring.
\Mod{Our main contribution lies in applying the hybrid approach to a practical ADS implementation and conducting experiments.}
More specifically, our contributions are as follows:
\begin{itemize}
  \item We adapt the {ground robot model} of Bohrer et al.~\cite{bohrerFormal2019} to abstract a trajectory-follower implementation of \Impl{Autoware} using parameters including ones that represent the accuracy of reaching waypoints (Sect.~\ref{s:modeling}). 
  Also, we re-prove the safety and liveness properties using \Impl{KeYmaera~X}; these properties ensure that the position and velocity of the vehicle remain within the assumed neighborhood of the planned trajectory (Sect.~\ref{s:verif:kyx}).
  \item We develop a runtime monitoring method to validate the conformance between the model and the \Impl{Autoware} implementation so that the verified properties implicitly apply to the latter (Sect.~\ref{s:verif:monitor}).
  The method instruments the \Impl{Autoware} controller node to extract the signals required for checking the assumption constraint of the model.
  The method is also implemented in a script for offline monitoring that evaluates each sub-formula of the constraint against a set of recorded data and aggregates the result.
  \item We conducted experiments to monitor \Impl{Autoware} executions using two simulation backends---a lightweight simulator (\Impl{simple\_planning\_simulator}) and a physics engine-integrated simulator (\Impl{AWSIM})---to validate the monitoring approach and to evaluate the waypoint-reachability accuracy achieved by the \Impl{Autoware} implementation (Sect.~\ref{s:experiment}).
  As a result, we confirmed (i)~the soundness of the monitoring method in terms of checking the reachability to waypoints, and (ii)~that the vehicle implementations reached the target with an accuracy of approximately $0.2$--$0.4$\,(m) in position and $0.8$\,(m/s) in velocity.
\end{itemize}

\section{Background}
\label{s:bg}

\Mod{As background, Sect.~\ref{s:autoware} introduces \Impl{Autoware}, Sect.~\ref{s:tf} describes the trajectory-following controller that we formalize and verify, and Sect.~\ref{s:modeling:dl} introduces the dL framework and \Impl{KeYmaera~X} underlying our verification method.}

\subsection{Autoware: Autonomous Driving Platform}
\label{s:autoware}

\Impl{Autoware}~\cite{katoAutoware2018}
\footnote{\url{https://autoware.org/}}
is an open-source software platform for autonomous driving, originally developed by TIER IV and maintained under the Autoware Foundation.
Built on ROS~2, it provides a full software stack that covers the major functional modules of an autonomous vehicle e.g. sensing, localization, perception, planning, and control.
Each module communicates through ROS~2 topics and services,
making the pipeline both modular and observable at runtime.
In this paper, we focus on the \emph{control}-related nodes, which are responsible for computing steering and acceleration commands that drive the vehicle along a planned trajectory represented by a set of waypoints.

\subsubsection{Pipeline}

\begin{figure}[t]
  \centering
  \includegraphics[width=0.85\textwidth]{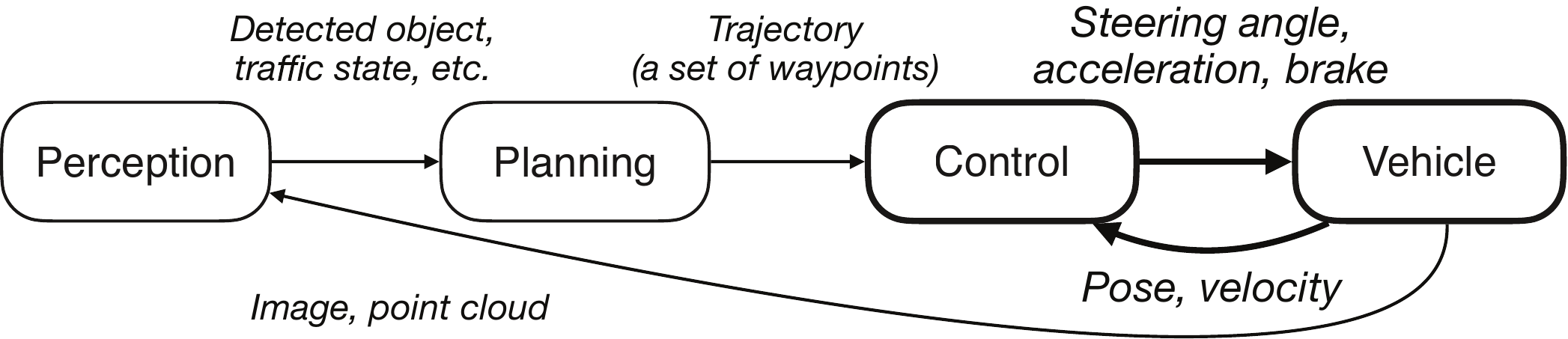}
  \caption{Pipeline overview of Autoware.}
  \label{f:pipeline}
\end{figure}

The \Impl{Autoware} software stack is organized as a pipeline (Fig.~\ref{f:pipeline}) of four functional stages:
\emph{perception}, \emph{planning}, \emph{control}, and \emph{vehicle}.
The perception component processes raw sensor data to produce a world model (detected objects, ego localization, etc.),
which the planning component uses to compute a reference trajectory consisting of waypoints toward the goal.
The control component then converts the trajectory into actuator commands---steering angle and acceleration---and sends them to the vehicle.
The vehicle component represents the physical interface, which can be a real automobile or a simulator
such as \Impl{AWSIM}\footnote{\url{https://github.com/autowarefoundation/AWSIM}} or \Impl{simple\_planning\_simulator} bundled with \Impl{Autoware}.
The vehicle state (position, heading and velocity) is fed back to the perception component, forming an outer feedback loop that closes the pipeline.
Within this outer loop, the control and vehicle components form an inner feedback loop, in which the vehicle state is continuously reported to the controller, which adjusts commands at each step to track the planned trajectory.

\subsubsection{Environment}
\label{s:env}

The operating environment of an autonomous vehicle is described by a \emph{scenario}, which specifies a map, a driving \emph{task}, and the behavior of surrounding agents such as other vehicles and pedestrians.
Each task contains additional parameters such as speed and time limits.
In this paper, we consider a simple planning task on a fixed map of Shinjuku, Tokyo. 
The task is defined by a start pose and a goal pose, both located on the road network.
The vehicle begins at rest and is required to reach the goal pose and come to a stop.

\subsection{Trajectory-Following Control}
\label{s:tf}

\begin{figure}[t]
  \centering
  \includegraphics[width=0.6\textwidth]{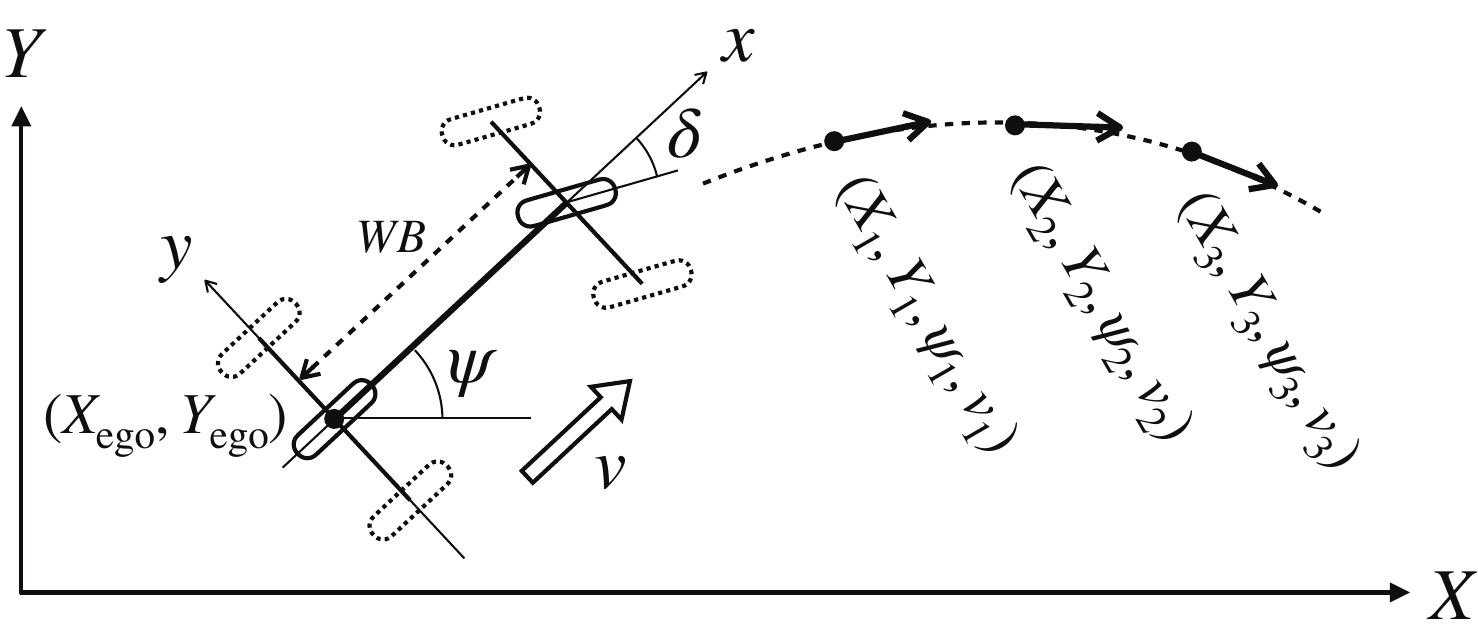}
  \caption{Bicycle model and objective trajectory.}
  \label{f:bicycle_traj}
\end{figure}

A controlled vehicle can be illustrated as shown in Fig.~\ref{f:bicycle_traj}.
Vehicles are considered to be a dynamical system with the state variable $(X,Y,\psi,v) \in \mathbb{R}^4$, where $(X,Y)$ represents a planar position (rear axle midpoint; shown as $(X_\mathrm{ego},Y_\mathrm{ego})$ in the figure) in a global coordinate system, $\psi$ is a planar yaw angle of the body rotation, and $v$ is a {longitudinal velocity}, and the input $u = (\delta,a)$, where $\delta$ is a steering angle and $a$ is an acceleration (we represent braking commands using negative accelerations).
The vehicle is also accompanied by local coordinate axes $(x,y)$, which we will use in our subsequent modeling.
In this work, we assume a \emph{kinematic bicycle model}; thus, the system is described by
\begin{equation} \label{e:tf_dyn}
  X'=v \cos(\psi), ~ Y'=v \sin(\psi), ~
  \psi'=\frac{v}{\mathit{WB}}\tan(\delta), ~
  v'=a,
\end{equation}
where $\mathit{WB}$ represents the wheel base.
In control, we refer to points in the workspace as \emph{waypoints} $W = (X,Y,\psi,v) \in \mathbb{R}^4$, each of which consists of a pose $(X,Y,\psi)$ and a velocity $v$.
Then, \emph{trajectories} are formulated as finite sequences of waypoints $W_1,\ldots, W_N$.

In a \emph{trajectory-following} task, an objective trajectory is given, and we control the ego vehicle to follow it.
At each step $j$ of the execution, the controller selects a target waypoint $(X_j,Y_j,\psi_j,v_j)$ from the trajectory and computes an input $(\delta_j,a_j)$ (we consider it as a step-wise constant signal) to make the state of the ego vehicle as close as possible to it after a certain period of time has elapsed. Namely, the target state will be the pose $(X_j,Y_j,\psi_j)$, and the velocity $v_j$.

\subsubsection{Trajectory Follower Node in Autoware}

The control functionality in \Impl{Autoware} is implemented as a ROS~2
\emph{composable node container} named \lstinline|/control/control_container|,
which hosts multiple nodes within a single OS process,
reducing inter-node communication overhead compared to separate processes.
%
The node 
\lstinline|/control/trajectory_follower/controller_node_exe|
inside the container implements the trajectory-following functionality.
It subscribes to a reference trajectory published by the planning module and outputs control inputs $(\delta,a)$ for the vehicle.
The computation of $\delta$ and $a$ is divided into two independent sub-problems handled
by plugin objects:
a \emph{lateral controller} that computes steering commands $\delta$ to follow the waypoint geometry, and a \emph{longitudinal controller} that computes
acceleration and deceleration commands $a$ to track the target velocity profile.
Two lateral controller implementations are provided:
\lstinline|MpcLateralController| based on a model predictive control (MPC) method 
and 
\lstinline|PurePursuitLateralController| based on a pure-pursuit method.
%
For longitudinal control,
\lstinline|PidLongitudinalController| provides a PID-based velocity controller.
In the experiment, we use the pure-pursuit lateral controller
combined with the PID longitudinal controller.

\subsection{Differential Dynamic Logic (dL) and KeYmaera~X}
\label{s:modeling:dl}

\emph{dL}~\cite{platzerLogical2018} is a logic for specifying and
verifying hybrid systems, in which discrete control and continuous dynamics
interact.
The syntax of hybrid programs $\alpha$ and dL formulas $\varphi$ is defined by the following
grammar, where $x$ is a real-valued variable whose value changes during execution, $\theta$ a real arithmetic term, 
and $Q$ a first-order formula over the reals.
\begin{align*}
  \alpha ~~::=~~~ &
    x := \theta
    \mid x := *
    \mid x' = f(x)\,\&\,Q
    \mid {?}Q
    \mid \alpha;\alpha
    \mid \alpha \cup \alpha
    \mid \alpha^* \\
  \varphi ~~::=~~~ &
    \theta \sim \theta
    \mid \lnot\varphi
    \mid \varphi \land \varphi
    \mid \varphi \lor \varphi
    \mid \varphi \to \varphi
    \mid \forall x\,\varphi
    \mid \exists x\,\varphi
    \mid [\alpha]\varphi
    \mid \langle\alpha\rangle\varphi
\end{align*}
A \emph{hybrid program} combines discrete assignments ($x := \theta$), nondeterministic assignments ($x := *$), and continuous evolution governed by a differential equation ($x' = f(x)\,\&\,Q$ within domain $Q$) using e.g. sequential composition ($\alpha;\alpha$) and repetition ($\alpha^*$).
\Mod{Atomic properties $\theta \sim \theta$ are comparisons on real arithmetic terms.}
Properties of a program are expressed using two modalities:
The \emph{box} formula $[\alpha]\varphi$ asserts that $\varphi$ holds after every possible execution of $\alpha$ (safety),
and the \emph{diamond} formula $\langle\alpha\rangle\varphi$ asserts that some execution of $\alpha$ reaches a state satisfying $\varphi$ (liveness).

In the next section, we model the waypoint-following vehicle as a hybrid program of the form $(\mathsf{ctrlX};\mathsf{plantX})^*$, where $\mathsf{ctrlX}$ represents one control step and $\mathsf{plantX}$ represents the continuous vehicle dynamics, and express its safety and liveness properties as dL formulas using the modalities.

As an example, we describe a system $\Ex$ in which a particle at $(x_\Ex, y_\Ex)$ rotates clockwise around the origin, possibly converging toward the origin (when $u_\Ex < 0$). 
Using the state variables $x_\Ex$, $y_\Ex$, and $u_\Ex$, we can model it as follows:
\begin{align*}
  \mathsf{ctrlE} ~\equiv~~
    & u_\mathsf{E} := *;~ ?u_\mathsf{E} \le 0 &
  \mathsf{plantE} ~\equiv~~ 
    & \{~ x_\mathsf{E}' = u_\mathsf{E} x_\mathsf{E} + y_\mathsf{E},~ y_\mathsf{E}' = -x_\mathsf{E} + u_\mathsf{E} y_\mathsf{E} ~\}
\end{align*}
$\mathsf{ctrlE}$ assigns an arbitrary real number to $u_\Ex$, and then the `$?$' operator blocks any assignment of a number greater than $0$.
$\mathsf{plantE}$ describes continuous behavior using a differential equation, where $u_\Ex$ is implicitly assumed to be a constant signal.
Let $\Init_\Ex$ be $x_\Ex=1 \land y_\Ex=0$. 
We can specify the safety property that the particle always remains within the disk of radius~1 by the dL formula 
$\Init_\Ex \to [(\mathsf{ctrlE}; \mathsf{plantE})*](x_\Ex^2 + y_\Ex^2 \le 1)$.
We can specify the liveness property that the particle eventually reaches the disk of radius~$\tfrac{1}{2}$ by
$\Init_\Ex \to \langle(\mathsf{ctrlE}; \mathsf{plantE})*\rangle(x_\Ex^2 + y_\Ex^2 \le \tfrac{1}{4})$.
The validity of these formulas is provable using the deduction system of dL.

\Impl{KeYmaera~X}\footnote{\url{https://keymaerax.org/}; \url{https://www.wolfram.com/engine/}\label{fn:kyx}} is an interactive theorem prover for hybrid systems based on dL.
It provides a deductive proof scheme based on the sequent calculus for dL, together with a set of dedicated proof tactics.
A particularly useful feature is the automatic tactics using Wolfram Engine\footnotemark[\getrefnumber{fn:kyx}]
such as \Impl{QE} to eliminate quantifiers in real arithmetic and \Impl{ODE} to solve differential equations.

\section{Modeling a Waypoint-Following Vehicle}
\label{s:modeling}

In this section, we model a vehicle and a trajectory-following controller as two variants $\mathsf{X} \in \{\TF, \GR\}$ of hybrid systems, in which a discrete controller $\mathsf{ctrlX}$ and continuous vehicle dynamics $\mathsf{plantX}$ alternate in a loop.
We formalize models and prove their properties in dL (Sect.~\ref{s:modeling:dl}).
We construct two models in Sect.~\ref{s:modeling:tf} and Sect.~\ref{s:modeling:gr}, expecting that the behaviors of the implementation are subsumed by those of the abstract models.
This inclusion relation is later confirmed empirically in the monitoring experiment.

\subsection{Trajectory Follower as a Hybrid Program}
\label{s:modeling:tf}

We first model the trajectory-following vehicle in Sect.~\ref{s:tf} as a hybrid program $\mathsf{TF}$, which describes the overall process implemented in \Impl{Autoware}, with details abstracted as black-box functions.
\Mod{This program is described using trigonometric functions and its properties are difficult to prove with \Impl{KeYmaera~X}; we therefore prove them after abstracting the program in Sect.~\ref{s:modeling:gr}.}

The $\mathsf{TF}$ model uses the variable vector $\mathcal{V}$ whose components include the following variables: the pose 
$(X\,\text{(m)}, Y\,\text{(m)}, \psi\,\text{(rad)})$, 
the longitudinal velocity $v\,\text{(m/s)}$ and acceleration $a\,\text{(m/s$^2$)}$, and the steering angle $\delta\,\text{(rad)}$.
Variables $a$ and $\delta$ are updated by the controller and remain constant otherwise.
The controller and plant are specified as follows:
\begin{align*}
  \mathsf{ctrlTF} ~\equiv~~
    & (X_{\mathrm{w}}, Y_{\mathrm{w}}, \psi_\mathrm{w}, v_\mathrm{w}) := *;~
      ?\mathit{Feas}_\TF; \\
    & \delta := \mathit{strCtrl}(\mathcal{V},X_\mathrm{w},Y_\mathrm{w});~
      a := \mathit{lngCtrl}(\mathcal{V},v_{\mathrm{w}}) \\
  \mathsf{plantTF} ~\equiv~~ 
    & t := 0;~ \{~ \text{Eq.~\eqref{e:tf_dyn}}, t'=1 ~\&~ t \le 
    T
    ~\}
\end{align*}

$\mathsf{ctrlTF}$ provides a high-level description that combines the processes of both the lateral and longitudinal controllers of \Impl{Autoware}. 
It mainly consists of three steps.
\textbf{(1)}~A target waypoint $(X_\mathrm{w}, Y_\mathrm{w}, \psi_\mathrm{w}, v_\mathrm{w})$ is selected. While these selected waypoints must satisfy a feasibility constraint $\mathit{Feas}_\TF$ that is implicitly assumed,
we treat it as a parameter and do not specify it explicitly. 
The waypoint retrieval process is performed independently in the lateral and longitudinal controllers; 
although different waypoints may be selected in each controller, we express them without loss of generality as a single tuple.
%
\textbf{(2)}~The steering angle $\delta$ is computed by the function $\mathit{strCtrl}()$ as $\arctan(\mathit{WB}\,k)$ where $k$ is the curvature of the arc connecting the pose of the ego vehicle to the target waypoint (the computation of $k$ will be described in the next section). 
\textbf{(3)}~The longitudinal acceleration $a$ is computed with $\mathit{lngCtrl}()$. Although the implementation basically follows PID control, 
it also incorporates velocity error computation with a filtering process, and differentiation using past values; 
here, we treat it as a black-box function and give no definition.

\Mod{$\mathsf{plantTF}$ describes the ordinary differential equation (ODE)~\eqref{e:tf_dyn} combined with a timer that restricts the evolution to at most a duration $T$. The default setting in the implementation is $T=0.03$.}

\subsection{Abstraction as a Ground Robot Model}
\label{s:modeling:gr}

\begin{wrapfigure}[12]{r}{.45\textwidth}
    \vspace*{-2.5em}
    \includegraphics[width=.45\textwidth]{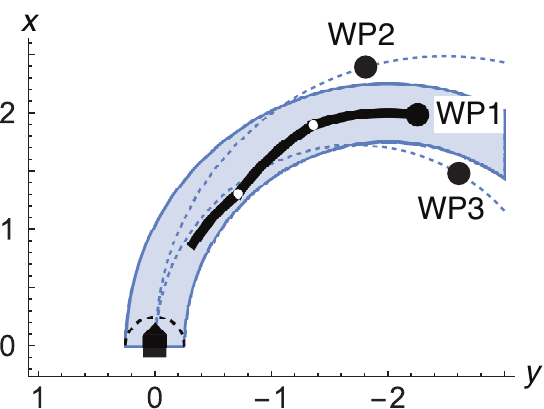} 
    \vspace*{-1.5em}
    \caption{Ground robot model.}
    \label{f:gr}
\end{wrapfigure}

\Mod{Abstraction plays a key role in proving properties of complex models with \Impl{KeYmaera~X}, as it allows its computer-algebra-based tactics to be applied successfully.
Verification also relies on tactics that require additional auxiliary expressions, such as differential variants, which are easier to find for simplified models.}
Here, we introduce a slightly modified version of the ground robot model proposed by Bohrer et al.~\cite{bohrerFormal2019}, which describes a robot following a target waypoint, and explain how it can be viewed as an abstraction of $\TF$.
We call this model $\GR$ in the sequel.
\Mod{Describing the dynamics in a local coordinate system eliminates the need for trigonometric functions, while replacing the detailed control law with a nondeterministic choice of the curvature and acceleration further simplifies the model.}

The $\GR$ model is based on a local coordinate system $(x,y)$ centered at the rear axle midpoint of the ego vehicle (as shown in Fig.~\ref{f:bicycle_traj}); here, we use only the local coordinates and describe how the target waypoint approaches the ego vehicle in relative terms, as shown in Fig.~\ref{f:gr}.
Following the convention in vehicle modeling, the vertical axis represents $x$ and the horizontal axis represents $y$, whose positive direction points to the left.
Tab.~\ref{t:vars} lists the variables and parameters used in the model.
%
The controller and plant are defined as follows:
\begin{align*}
  \mathsf{ctrlGR} ~\equiv~~
    & k := *;~ ?\mathit{Feas};~ a := *;~ ?\mathit{Go} \\
  \mathsf{plantGR} ~\equiv~~ 
    & t := 0;~ \{~ x' = v (k y - 1),~ y' = -v k x,~ v' = a,~ t' = 1 \\
    & \hspace{4em} ~\&~ t \le T \land v \ge 0 ~\}
\end{align*}

The control objective is to set $(k,a)$, where $k$ represents the curvature of the trajectory and $a$ is an acceleration within the interval $[-B,A]$, so as to bring the relative waypoint position $(x,y)$ into the $\Eps$-ball around the origin and to keep the velocity within $[v_l,v_h]$.
The nondeterministic process of $\mathsf{ctrlGR}$ includes the case where the value of $k$ is perturbed;
it also represents speed control with the admissible values of $a$, which encompasses the case of PID control.
The differential equation of $\mathsf{plantGR}$ describes the behavior in which the waypoint moves toward the ego vehicle's position along an arc of curvature $k$. 

\begin{table}[!t]
    \centering
    \scriptsize
    \caption{Variables (left) and constant parameters (right) of $\GR$.} \label{t:vars}
    \setlength{\tabcolsep}{.5em}
    \begin{tabular}{@{}cl|cl@{}}
        \toprule
        Var. (unit) & Meaning & Param. & Meaning \\
        \midrule
        $(x\,\text{(m)},y\,\text{(m)})$ & Relative position of the target. & 
        $\Eps$ & Bound on positional errors. \\
        $k\,\text{(m$^{-1}$)}$ & A controlled curvature. & 
        $v_l, v_h$ & Bounds on the target velocity. \\
        $v\,\text{(m/s)}, a\,\text{(m/s$^2$)}$ & Longitudinal velocity and acceleration. & 
        $A, B$ & Bounds on accelerations. \\
        $t\,\text{(s)}$ & Elapsed time of a continuous step. & 
        $T$ & Maximum step time. \\
        \bottomrule
    \end{tabular}
\end{table}

In trajectory following, reaching a target waypoint typically requires multiple steps. Unlike $\TF$, the $\GR$ model focuses on a single target and describes the process of achieving it over multiple steps. Accordingly, the target position $(x,y)$ is given as an arbitrary initial value at time 0, 
and the target velocity is represented by a constant interval parameter $[v_l,v_h]$ (where $v_l \ge 0$ and $v_l < v_h$).
The target $(x,y)$ is shown as ``WP1'' in Fig.~\ref{f:gr}.
In the verification, we assume an initial condition $\mathit{Init}$ specified as follows (each sub-constraint is explained below; see Appendix~\ref{a:constraint} for the complete specifications):
\begin{align*}
  \mathit{Init} ~~\equiv~~&
  \multispan{3}{$T>0 ~\land~ \Eps>0 ~\land~ A>0 ~\land~ B>0 ~\land~ \mathit{Inv}$ \hfil} \\
  \mathit{Inv} ~~\equiv~~&
  v \geq 0 ~\land~ \mathit{Feas} ~\land~ \mathit{GoA} ~\land~ \mathit{GoB} \quad
  & \mathit{Feas} ~~\equiv~~& \mathit{Ann} ~\land~ \mathit{Speed} \\
  \mathit{Ann} ~~\equiv~~&
  \multispan{3}{$
  |k| \Eps \leq 1 
  ~\land~
  |k (x^2 + y^2 - \Eps^2) / 2 - y| < \Eps$ \hfil} \\
  \mathit{Speed} ~~\equiv~~&
  \multispan{3}{$0 \leq v_l < v_h ~\land~ A T \leq v_h - v_l ~\land~ B T \leq v_h - v_l$ \hfil} \\
  \mathit{GoA} ~~\equiv~~&
  \multispan{3}{$v_l \leq v ~\lor~ (1 + |k|\Eps)^2 \left(v T + \tfrac{a}{2} T^2 + \tfrac{v_l^2 - (v + a T)^2}{2 A}\right) + \Eps \leq x$ \hfil}
  %
\end{align*}

\Mod{The $\GR$ model is an abstraction that keeps targeting a single waypoint over multiple steps, whereas pure-pursuit control may switch the waypoints being followed based on the lookahead distance.}
A key point of the verification is how other waypoints appearing at intermediate steps affect the accuracy of reaching the initial target. 
\Mod{Fig.~\ref{f:gr} depicts two intermediate waypoints ``WP2'' and ``WP3'' that influence the curvature of the trajectory;
the white dots indicate where the trajectory's curvature changes.}
In the model, intermediate waypoints are not explicitly described, but we assume that they influence the control input $(k,a)$ computed at each step.
We assume that the waypoints appear in the vicinity of the initial target and describe admissible conditions by $\mathit{Feas}$ (specified above) and $\mathit{Go}$, which is specified as:
\begin{equation*}
  \mathit{Go} ~~\equiv~~
  -B \leq a \leq A ~\land~ v + a T \geq 0 ~\land~ \mathit{GoA}' ~\land~ \mathit{GoB}'
\end{equation*}

The constraint $\mathit{Ann}$ states that the vehicle and the waypoint are within the $\Eps$-annulus (an arc surrounded by the margin $\Eps$).
In pure-pursuit control, the curvature is computed as $k := \tfrac{2y}{x^2+y^2}$, which defines the arc connecting the vehicle and the waypoint;
$\mathit{Feas}$ is formulated by incorporating multiple $\Eps$ terms to represent an annulus around the arc.
$\mathit{Speed}$ states that the bounds $v_l$ and $v_h$ are non-negative; it also states their ordering, and the relationship between the error $v_h-v_l$ and the maximum speed change over the duration $T$.
Basic constraints on the velocity and acceleration are prepared as $\mathit{GoA}$ and $\mathit{GoB}$; 
$\mathit{GoB}$ is defined analogously to $\mathit{GoA}$.
Their conjunction requires that either the speed is already within $[v_l,v_h]$, or there is sufficient longitudinal distance to accelerate (with $A$) or decelerate (with $-B$) to the target within $T$.
$\mathit{Go}$ specifies basic conditions on $a$ and $v$ appended with $\mathit{GoA}'$ and $\mathit{GoB}'$, which are $\mathit{GoA}$ and $\mathit{GoB}$ slightly modified for ease of verification.

Model $\GR$ describes only the local behaviors of $\TF$.
Given a pose $(X,Y,\psi)$ and a target position $(X_\mathrm{w},Y_\mathrm{w})$ in $\TF$, the state of $\GR$ is uniquely determined as
$(x,y)^\top := R_{-\psi} ((X_\mathrm{w},Y_\mathrm{w}) - (X,Y))^\top$ where $R_{-\psi}$ is the rotation matrix by $-\psi$.
We can consider a simulation relation between $\TF$ and $\GR$.
\Mod{For executions in which the premise constraints are confirmed by runtime monitoring, the following proposition provides the basis for transferring the safety property, proved later, to the behavior of the implementation.}
\begin{proposition} \label{prop}
  Assume constants in Tab.~\ref{t:vars}.
  Consider the state equivalence relation between $\TF$ and $\GR$ with the coordinate transformation based on the first waypoint $(X_{\mathrm{w}},Y_{\mathrm{w}},v_{\mathrm{w}})$ that appears in an execution of $\TF$.
  If the coordinate-transformed initial state of $\TF$ satisfies $\mathit{Init}$, and intermediate assignments of $k$ and $a$ are not blocked by $\mathit{Feas}$ and $\mathit{Go}$,
  then $\GR$ simulates $\TF$.
\end{proposition}
The proof is given in Appendix~\ref{a:proof}.

\section{Hybrid Verification Process}
\label{s:verif}

In this section, we describe a verification process for trajectory-following vehicles, modeled as $\GR$ and implemented with \Impl{Autoware}.
Following the existing verification~\cite{bohrerFormal2019}, we first verify a \emph{safety} and a \emph{liveness} property of the $\GR$ model:

  The \emph{safety property} below states that throughout every execution starting from $\Init$, the invariant $\mathit{Inv}$ is maintained. In particular, the vehicle always remains within the $\Eps$-annulus formed by the ego position and the target waypoint, and whenever the vehicle reaches the $\Eps$-ball of the target, its velocity lies within the interval $[v_l, v_h]$.
  \begin{equation*} \label{e:safety}
      \mathit{Init} ~\to~
      [\{\mathsf{ctrlGR};~ \mathsf{plantGR}\}*]
      (\mathit{Inv} ~\land~ (|\!|(x,y)|\!| \leq \Eps \to v \in [v_l,v_h]))
  \end{equation*}
  
  The \emph{liveness property} considers any reachable state in which the additional premises $v>0$ and $x>0$ hold;
  by the safety property above, any such state lies within the annulus. 
  Then, there exists a control sequence that eventually brings the ego vehicle into the $\Eps$-ball around the target with velocity in $[v_l, v_h]$.
  \begin{multline*} \label{e:liveness}
      %
      \mathit{Init} ~\to~
      [\{\mathsf{ctrlGR};~ \mathsf{plantGR}\}*] \bigl( (v>0 ~\land~ x>0) ~\to~ \\
      \langle\{\mathsf{ctrlGR}_a;~ \mathsf{plantGR}\}*\rangle ( |\!|(x,y)|\!| \le \Eps ~\land~ v \in [v_l,v_h]) \bigr),
  \end{multline*}
  \Mod{where $\mathsf{ctrlGR}_a$ is a variant in which the assignment to the curvature $k$ is removed. The diamond fragment therefore represents the phase in which no other waypoint appears before the target is reached.}
%
The additional premises in the liveness property assert that the vehicle is advancing toward waypoints that lie ahead and are not restrictive in practice.

Our verification follows a \emph{hybrid} approach that combines formal model verification and runtime verification:
\textbf{(1)}~We first conduct a mechanical proof using \Impl{KeYmaera~X} to confirm that the abstract $\GR$ model satisfies both properties (Sect.~\ref{s:verif:kyx}).
%
\textbf{(2)}~We then monitor the \Impl{Autoware} execution by collecting data from the controller implementation, which, together with a vehicle in a simulator, forms the feedback loop 
(Sect.~\ref{s:verif:monitor}).
More precisely, we monitor the satisfaction of the constraints $\Init$, $\mathit{Feas}$, and $\mathit{Go}$ across comprehensive simulation runs of \Impl{Autoware}, examining the compatibility of the implementation with the $\GR$ model.

\subsection{Mechanical Proof with $\mathsf{KeYmaera~X}$}
\label{s:verif:kyx}

The original proofs for the $\GR$ model~\cite{bohrerFormal2019} were developed in an earlier version of \Impl{KeYmaera~X} and rely on several tactics that are no longer available in recent releases.
We proved the properties from scratch using \Impl{KeYmaera~X}~5.1.2, following the same overall structure but replacing the unavailable tactics with standard ones; the only non-standard tactic we needed was \Impl{mond} (available in the current release but hidden from the user interface), the monotonicity rule of the diamond modality.
Proof scripts are available at: \url{https://github.com/cpcp-lab/KeYmaeraX-projects}

\paragraph{Safety.}
The safety proof proceeds by showing that the invariant $\mathit{Inv}$ is a loop invariant of $(\mathsf{ctrlGR};\mathsf{plantGR})^*$.
For the continuous step, we verify that each conjunct of $\mathit{Inv}$ is preserved by $\mathsf{plantGR}$.
This reduces to showing that each sub-formula of $\mathit{Inv}$ can be established as a \emph{differential invariant} of the ODE; the tactic \Impl{dI} automates this check and suffices for most conjuncts after sufficient case splitting.
The conjuncts $\mathit{GoA}$ and $\mathit{GoB}$ require additional case splits on the sign of $a$, followed by algebraic manipulations.
Absolute values appearing in $\mathit{Ann}$ are handled by the case-splitting process.

\paragraph{Liveness.}
The proof shows that a control strategy drives the vehicle to the target.
We split the proof according to the initial longitudinal velocity $v$ into three cases: (i)~$v < v_l$, (ii)~$v_l \leq v \leq v_h$, and (iii)~$v > v_h$.
In cases~(i) and~(iii), we first prove that the velocity can be brought into $[v_l, v_h]$ by choosing $a = A$ or $a = -B$, respectively, and then apply \Impl{mond} to chain the resulting reachability into that of case~(ii).
In case~(ii), a key step is the use of the \Impl{dV} tactic
that shows that the value $2 v_0 x_t$ is decreasing throughout the continuous step, where $v_0$ is the value of $v$ at the start of the step and $x_t$ is an auxiliary variable such that $0<x_t<x$;
the last condition requires the use of the \Impl{barrier} tactic.

\subsection{Runtime Monitoring of the Constraints}
\label{s:verif:monitor}

We have modified the \Impl{Autoware} toolchain to monitor the simulated executions to evaluate their conformance with the formally verified $\GR$ model.


\paragraph{Instrumentation.}
We instrument the \Impl{PurePursuitLateralController} class of \Impl{Autoware Universe}~0.50.0\footnote{\url{https://github.com/cpcp-lab/autoware_universe}} to publish the data needed for monitoring.
It computes a target position (the intersection of the lookahead circle with the trajectory) by interpolation but does not expose the associated waypoint velocity.
We extend it to select the first waypoint beyond the lookahead distance and publish its content $(X_\W,Y_\W,v_\W)$, together with the current state $(X,Y,\psi,v)$ of the ego vehicle.
%
In monitoring, we also use an existing topic of the controller node \inlst{/control/command/control_cmd} to obtain the steering angle $\delta$ and longitudinal acceleration~$a$ computed by the controller.

\paragraph{Automated simulation with \Impl{scenario\_simulator\_v2}.}
To orchestrate the simulation process and to collect log data comprehensively, we use our own modified version of \Impl{scenario\_simulator\_v2},%
\footnote{\url{https://github.com/cpcp-lab/scenario_simulator_v2}}
an \Impl{Autoware}-integrated framework for automated scenario-based simulation.
%
In our experiments, we used the \Impl{scenario\_test\_runner} script, which executes \Impl{Autoware} together with a backend vehicle simulator following an input scenario description.
As a vehicle simulator, one can use either:
\Impl{Simple} (\Impl{simple\_planning\_simulator}), a built-in lightweight implementation based on a kinematic bicycle model and ROS visualization, or 
\Impl{AWSIM}~2.0.0, an external program that performs more detailed vehicle simulation using the physics engine provided by {Unity}.

We modified \Impl{scenario\_simulator\_v2} to configure a set of driving tasks on the Shinjuku map; each task specifies a start pose and drives the ego vehicle toward a reachable goal pose within a specified effort range.
We also added a logging function that subscribes to the two kinds of data from the controller node (the published waypoint and vehicle state, and the control commands) and saves them into separate log files. 


\paragraph{Offline monitoring script.}
We implement an offline monitoring script\footnote{\url{https://github.com/cpcp-lab/autoware_control_monitor}} in Python that processes the recorded logs as follows.
\textbf{(1)}~Stepwise elements in the two log files are synchronized by timestamp.
\textbf{(2)}~From each synchronized trace (a sequence of elements), possibly overlapping \emph{windows} are extracted, one starting at every 50th step; each window starts from an initial pose of the ego vehicle and extends until the vehicle reaches the initial target waypoint, so it also covers the intermediate waypoints visited along the way.
\textbf{(3)}~For each window, the script converts each position coordinate to local coordinates and evaluates:
\begin{enumerate}[label=\Alph*.]
  \item Whether the initial state of the window satisfies $\mathit{Init}$.
  \item Whether the ego vehicle eventually reaches the $\Eps$-ball of the waypoint and the velocity falls within the target interval $[v_l,v_h]$. 
  Here, we represent the interval as $v_\W + [-\Eps_v,\Eps_v]$ where $v_\W$ is the velocity given by the target waypoint and $\Eps_v$ is a positive real number specified in the experiment.
  When verifying the reachability, per-step checking may miss an intersection that occurs in-between;
  therefore, we linearly interpolate between consecutive samples and check whether any interpolated segment intersects the $\Eps$-ball. If $x$ becomes negative without such a crossing, the window is classified as a non-reaching case.
  \item Whether any step of the window violates the monitoring conditions $\mathit{Feas}$ and $\mathit{Go}$.
\end{enumerate}
The results are aggregated over all windows and reported as the fractions of windows that (a)~satisfy $\mathit{Init}$, (b)~reach the target, and (c)~violate each constraint.

\Mod{In this work, monitoring is performed offline; however, it is lightweight enough to run online. Using online monitoring for safety enforcement is left as future work.}

\section{Experiment}
\label{s:experiment}

\Mod{The goal of this paper is to transfer the safety and liveness guarantees proved for the abstract $\GR$ model (Sect.~\ref{s:verif:kyx}) to the actual \Impl{Autoware} implementation through runtime monitoring (Sect.~\ref{s:verif:monitor}), which bridges the gap between them.}
To this end, we have conducted experiments to evaluate \Mod{the monitoring process of} the hybrid verification method, addressing the following questions:
\textbf{(RQ1)}~Does runtime monitoring (of a \emph{valid} window, i.e., one satisfying all the monitoring constraints) guarantee that the vehicle reaches a neighborhood of the target waypoint? This question checks whether monitoring soundly reflects the guarantees proved for the model.
\textbf{(RQ2)}~Under which accuracy parameters $\Eps$ and $\Eps_v$ do the actual execution results of \Impl{Autoware} conform well to the monitoring constraints? This question quantifies the accuracy with which the implementation's behavior is subsumed by the verified model.
\textbf{(RQ3)}~Do the results generalize to the more accurate vehicle model provided by \Impl{AWSIM}? This question addresses the external validity of the above findings beyond a single simulation backend.

\Mod{The implementation and the experimental results are described in Appendix~\ref{a:artefact} and available at:
\url{https://doi.org/10.5281/zenodo.21510141}}

\subsection{Setup}

Experiments were conducted on a machine with a 4.5\,GHz AMD Ryzen~9 7950X processor, 64\,GB RAM, and an NVIDIA RTX\,5060\,Ti GPU\,(16\,GB VRAM).
\Impl{Autoware Universe}~0.50.0 was used.
\Mod{We mainly used the default values for the control parameter settings; verifying other settings is left as future work.}
The control step duration was clamped to $[0.015, 0.06]$\,(s), 
the acceleration inputs from the controller were clamped to $[-1, 1]$\,($\mathrm{m}/\mathrm{s}^2$),
and the PID parameters of the longitudinal controller were $k_p=1$, $k_i=0.1$, $k_d=0$.

\paragraph{Tasks.}
We randomly sampled 100~lanelets from the Shinjuku map.
For each, we enumerated all goal lanelets reachable within an effort range of $[10, 50]$, yielding a set of start--goal pairs.
For each pair, a driving task was configured and a simulation run was executed with each of the \Impl{Simple} and \Impl{AWSIM} backends.
Both initial and goal poses were placed at the centerline start of the selected lanelets; also, the initial and final velocities were set to zero.

\paragraph{Monitoring parameters.}
The offline monitoring script was applied to the recorded logs with the following parameters.
We fixed $A = B = 1.5$\,($\mathrm{m}/\mathrm{s}^2$) to accommodate most of the outliers occasionally present in the logs that exceed the output clamp bounds of the controller implementation.
We also fixed $T = 0.03$\,(s), the upper bound of the control step duration we observed, which lies within the clamp interval of the implementation.
We evaluated over a grid of accuracy parameters: $\Eps \in \{0.025,0.05, 0.1, 0.2, 0.4, 0.8, 1.6\}$\,(m) and $\Eps_v \in \{0.1, 0.2, 0.4, 0.8, \AB 1.6, \AB 3.2\}$\,(m/s).

\subsection{Results}

\begin{figure}[!t] \centering
  \includegraphics[width=0.375\textwidth]{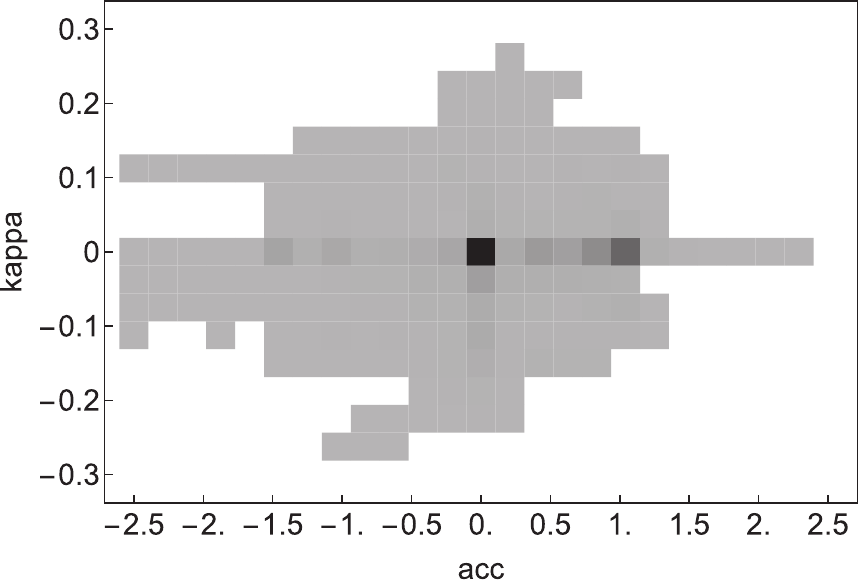}
  \hspace{2em}
  \includegraphics[width=0.375\textwidth]{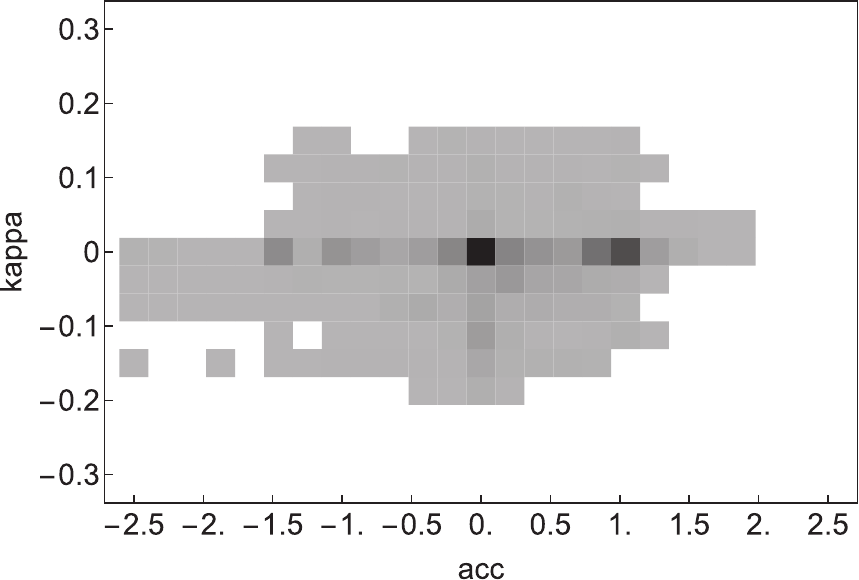}
  \vspace{-.5em}
  \caption{Distribution of control inputs $(a,k)$ (left: \Impl{Simple}, right: \Impl{AWSIM}).}
  \label{f:aks}
\end{figure}

We ran $163$ tasks with \Impl{Simple} and \Impl{AWSIM}. 
\Mod{The generated tasks are routes of diverse shapes;
the diversity of routes can be characterized by the required acceleration and steering (curvature) inputs (Fig.~\ref{f:aks}).}
One vehicle position from every recorded 50~steps was selected as the initial position of a window, and the process of reaching the target waypoint from each was monitored.
As a result, $2651$--$2817$ and $1425$--$1475$ windows were analyzed for \Impl{Simple} and \Impl{AWSIM}, respectively, for every accuracy parameter setting $(\Eps,\Eps_v)$.
For $7$ and $60$ of the $163$ tasks, 
respectively, no log files were generated due to planning or control failures.
For each accuracy setting, a number of windows were excluded in preprocessing because they contained irrelevant data after achieving the goal, or they did not cover the full trajectory to reach the target $\Eps$-ball.


Fig.~\ref{f:aks} shows the control inputs $(a,k)$ at each monitored sample, aggregated over a grid of values.
The set of generated tasks covered the admissible range $[-B,A] = [-1.5,1.5]$ of acceleration; outliers of up to $\pm 2.5$, which violate $\mathit{Go}$, also occurred with some frequency. They also covered the typical curvature range $[-0.2,0.2]$ for urban driving. The range observed in \Impl{Simple} was wider than that in \Impl{AWSIM}.

\begin{figure}[!t] \centering
  \includegraphics[width=0.85\textwidth]{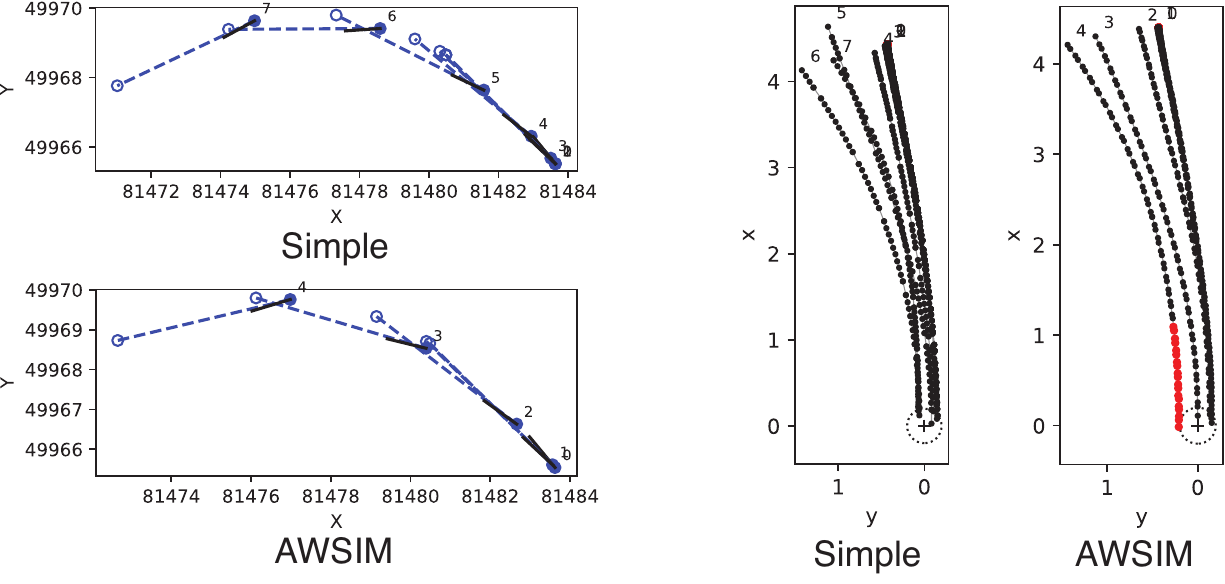}
  \caption{Example positional data. The left figures represent the tasks; filled and open circles denote initial and goal positions, respectively. The right figures represent the vehicle positions in the local coordinate system; black and red dots indicate vehicle positions satisfying and violating the monitoring constraints, respectively, with the setting $(\Eps, \Eps_v) = (0.2,0.8)$.}
  \label{f:trajs}
\end{figure}

Fig.~\ref{f:trajs} illustrates an example of the experimental process.
The left figures show the 7 and 4 windows executed when the same task was given to \Impl{Simple} and \Impl{AWSIM}, respectively.
Note that \Impl{AWSIM} failed before reaching the task's goal position.
The right figures show the monitoring results for each window based on the local coordinate system of the $\GR$ model. 

For each accuracy setting $(\Eps, \Eps_v)$, we aggregated the resulting status of each window.
First, windows in which the initial condition $\Init$ is not satisfied are grouped---denoted $\neg I$ (not initialized)---and excluded from further classification.
Next, the remaining windows are classified according to whether all monitoring constraints ($\mathit{Feas}$ and $\mathit{Go}$ at each step) are satisfied---denoted $V$ (valid)---and whether the vehicle reaches the $\Eps$-ball (and the $\Eps_v$-interval for the velocity) of the target waypoint---denoted $R$ (reached).

Representative results for selected $(\Eps, \Eps_v)$ values are shown in Tab.~\ref{t:xp} for both backends.
The table uses the accuracy $(0.2,0.8)$ as a baseline, varying each parameter from there.
Each column shows the percentage of windows in a given group relative to the total;
for example, the column ``$V \land R$'' corresponds to windows that satisfy the initial condition, all the monitoring constraints, and also reach the $\Eps$-ball.
Notably, for all tested configurations and both backends, no window fell into the group $V\land\neg R$ in any run.
%

\begin{figure}[!t] \centering
  \includegraphics[width=0.4\textwidth]{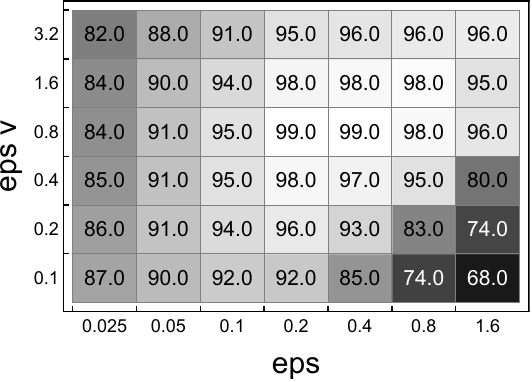}
  \hspace{1em}
  \includegraphics[width=0.4\textwidth]{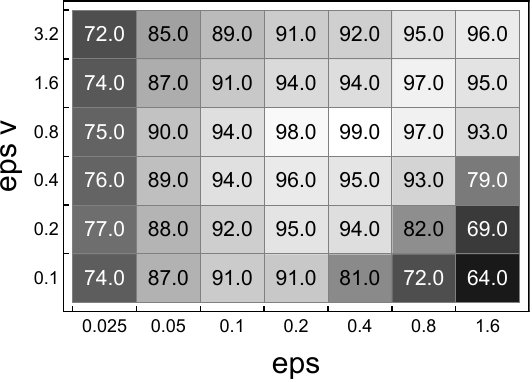}
  \vspace{-.5em}
  \caption{Heat map of true reachability judgment rates (left: \Impl{Simple}, right: \Impl{AWSIM}).}
  \label{f:hm}
\end{figure}

To analyze how the judgment outcome responds to the configured accuracy parameters, Fig.~\ref{f:hm} visualizes the \emph{true reachability judgment rate}, i.e., $\tfrac{\#(V \land R)}{\# R}$, for every parameter setting.

\begin{table}[!t]
    \centering
    \scriptsize
    \caption{Selected classification percentages of experimental results.} \label{t:xp}
    \setlength{\tabcolsep}{.5em}
    \begin{tabular}{@{}rr|rrrrr|rrrrr@{}}
        \toprule
        && \multicolumn{5}{c|}{\scriptsize \Impl{Simple}} & \multicolumn{5}{c}{\scriptsize \Impl{AWSIM}} \\
        \cmidrule(l){3-7} \cmidrule(l){8-12}
        $\Eps$ & $\Eps_v$ & $\neg I$ & $V \!\land\! R$ & $\neg V \!\land\! R$ & $V \!\land\! \neg R$ & $\neg V \!\land\! \neg R$ 
        & $\neg I$ & $V \!\land\! R$ & $\neg V \!\land\! R$ & $V \!\land\! \neg R$ & $\neg V \!\land\! \neg R$ \\
        \midrule
        0.2  & 0.8 & $ 9.4\%$ & $83.9\%$ & $ 1.1\%$ & $0.0\%$ & $ 5.7\%$
                   & $11.2\%$ & $76.2\%$ & $ 1.3\%$ & $0.0\%$ & $11.3\%$ \\
        0.05 & 0.8 & $ 9.5\%$ & $63.9\%$ & $ 6.6\%$ & $0.0\%$ & $20.1\%$
                   & $13.0\%$ & $38.3\%$ & $ 4.3\%$ & $0.0\%$ & $44.5\%$ \\
        0.8  & 0.8 & $ 9.5\%$ & $84.9\%$ & $ 1.8\%$ & $0.0\%$ & $ 3.8\%$
                   & $11.6\%$ & $79.4\%$ & $ 2.4\%$ & $0.0\%$ & $ 6.5\%$ \\
        0.2  & 0.2 & $ 9.5\%$ & $70.4\%$ & $ 3.0\%$ & $0.0\%$ & $17.2\%$
                   & $11.3\%$ & $63.8\%$ & $ 3.2\%$ & $0.0\%$ & $21.7\%$ \\
        0.2  & 3.2 & $ 9.4\%$ & $84.6\%$ & $ 4.5\%$ & $0.0\%$ & $ 1.5\%$
                   & $11.1\%$ & $76.7\%$ & $ 7.2\%$ & $0.0\%$ & $ 4.9\%$ \\
        %
        \bottomrule
    \end{tabular}
\end{table}

\subsection{Discussion}

\Mod{The zero count of the $V\land\neg R$ case for all execution windows shows that whenever a window satisfies the monitoring constraints ($V$), the vehicle indeed reaches the target waypoint ($R$);} that is, monitoring provides a \emph{sound} reachability judgment ($V \Rightarrow R$), answering \textbf{RQ1} positively.

The trajectory of the vehicle is influenced by intermediate waypoints as shown in Fig.~\ref{f:gr}. 
We therefore observe, from Tab.~\ref{t:xp} and the heat map in Fig.~\ref{f:hm}, how the conformance between the monitoring constraints and reachability changes as the accuracy parameter values are varied.
In Fig.~\ref{f:hm}, the settings with $\Eps \in \{0.2,0.4\}$ and $\Eps_v = 0.8$ achieve better judgment rates than the neighboring cells.
As the parameter values are reduced, the number of windows that reach the target also decreases.
When $\Eps$ is reduced, the $\mathit{Ann}$ constraint becomes tighter, and we have indeed observed states that fail to satisfy it. Similarly, when $\Eps_v$ is reduced, we observed multiple states that violate the $\mathit{Go}$ constraint. We believe that these violations reflect a trade-off between $\mathit{Ann}$ and $\mathit{Go}$ caused by the interaction between $\Eps$ and $\Eps_v$, i.e., by how close each of them is set to the actual accuracy of the implementation.
As the parameter values are increased, the number of unreachable cases decreases; it reaches zero when $\Eps \geq 1.6$ and $\Eps_v \geq 1.6$. Under these settings, violations of the monitoring constraints become prominent. We confirmed that violations of $\mathit{Go}$ are frequent.
An interaction between $\Eps$ and $\Eps_v$ is visible. Since $\mathit{Go}$ checks that the position and the velocity are feasible, parameter settings near the diagonal appear to yield higher constraint satisfaction.
Overall, we consider that the settings with $\Eps \in \{0.2,0.4\}$ and $\Eps_v = 0.8$ are neither too strict nor too loose, and conform well to the actual accuracy of the implementation (\textbf{RQ2}).

Regarding \textbf{RQ3}, the monitoring results for \Impl{AWSIM} show a similar trend, but with lower accuracy; namely, the number of windows that reach the $\Eps$-ball decreases, and the true reachability judgment rate also decreases.
For \Impl{Simple}, the reachability judgment rate exceeds $80\%$ for most settings with $\Eps \geq 0.2$ and $\Eps_v \geq 0.4$, whereas for \Impl{AWSIM}, this threshold is reached only when $\Eps \geq 0.4$ and $\Eps_v \geq 0.8$.
As can be seen in Fig.~\ref{f:trajs}, \Impl{AWSIM} tends to achieve lower waypoint-reaching accuracy than \Impl{Simple}.
Also, a number of tasks we fed to \Impl{AWSIM} resulted in failures.
Although the result logs contain error messages related to point cloud data and planning trajectories, the exact cause of these failures is unclear. 
As shown in Fig.~\ref{f:aks}, executions tended to fail when the trajectory curvature fell outside the range $[-0.2, 0.2]$. 
Investigating the root cause is left as future work.
\Mod{Since failed tasks were excluded from the analysis, we believe this does not threaten the internal validity of the above discussion.}


%

\section{Related Work}
\label{s:related}

A survey by Luckcuck et al.~\cite{luckcuckFormal2020} has systematically categorized formal specification and verification of autonomous and robotic systems.
Mehdipour et al.~\cite{mehdipourFormal2023a} specifically survey formal methods for ensuring road-rule compliance in autonomous driving.
These surveys confirm that while the field has advanced toward increasingly detailed formal models and toward implementation-level analysis, few works apply formal verification to the controller-implementation level of a practical ADS.
Our work addresses this gap by applying a hybrid approach to \Impl{Autoware}.

\emph{dL-based verification and monitoring.}
A line of work has applied dL and \Impl{KeYmaera~X} to autonomous vehicles.
Loos et al.~\cite{loosAdaptive2011} verified adaptive cruise control (ACC).
Mitsch and Platzer~\cite{mitschModelPlex2016} proposed \emph{ModelPlex}, which synthesizes provably correct runtime monitors from dL proofs; our monitor embedded in \Impl{Autoware} is similarly derived but implemented manually.
Mitsch et al.~\cite{mitschFormal2017} formally verified a variant of the $\GR$ model for obstacle avoidance and navigation tasks, with a runtime monitoring experiment.
Bohrer et al.~\cite{bohrerFormal2019} proved safety and liveness for the $\GR$ model and monitored the constraints against a simple vehicle implementation in a simulator; our $\GR$ model is slightly modified from theirs, and we extend the evaluation to the \Impl{Autoware} implementation and its dedicated simulators.
Lin et al.~\cite{linSafe2022} built a safety control filter on the same model and evaluated it in a racing simulator.
Prebet et al.~\cite{prebetVerification2025} applied ModelPlex to synthesize a safeguard for a neural network controller of an autonomous vehicle.

\emph{Formal verification of ADS.}
Rizaldi et al.~\cite{rizaldiFormalising2017} formalized traffic rules using abstract road and vehicle models in Isabelle/HOL and obtained monitors via code generation.
Nenchev et al.~\cite{nenchevCodeLevel2025} verified safety properties of an ACC implementation at the code level using CBMC.
Fritzsch et al.~\cite{fritzschExperiences2021} reported experiences with NuSMV for model checking an automotive control system.
These approaches treat the system as a discrete-time transition system; in contrast, our work models the continuous-time dynamics of vehicle control as a hybrid system and uses the resulting formal constraints to monitor the actual implementation.

\emph{Scenario-based testing.}
Testing with scenario-based simulations is a practical and popular approach for ADS.
Tang et al.~\cite{tangCollision2021} proposed a method that generates and tests diverse scenarios based on map topology.
Viswanadha et al.~\cite{viswanadhaAddressing2021} proposed a method based on abstract scenarios and falsification of safety properties.
Zhou et al.~\cite{zhouSpecificationbased2023} proposed a fuzzing-based method to search for test cases violating user-defined safety assertions.
Our method uses simple scenario configurations to test the basic control functions of an ADS. Analyzing corner-case scenarios remains future work.

\emph{Runtime verification.}
Huang et al.~\cite{huangROSRV2014} and Ferrando et al.~\cite{ferrandoROSMonitoring2020a} proposed ROSRV and ROSMonitoring, respectively, which are runtime verification tools that insert a monitoring node into ROS to enforce safety and security policies.
Nilsson et al.~\cite{nilssonCorrectbyConstruction2016} developed a correct-by-construction approach for ACC in which a monitor-based filter enforces safety at runtime.
Lin et al.~\cite{linReachFlow2020} proposed ReachFlow, which assures waypoint-following safety for self-driving cars online using reachability analysis.
Our work is distinguished by targeting the controller node internals of a full ADS stack and by evaluating both positional and velocity reachability with accuracy parameters.

\emph{Hybrid verification approaches.}
Fisher et al.~\cite{fisherVerifying2013} proposed verifying high-level agent decisions by model checking while testing lower-level implementations through simulation.
Aniculaesei et al.~\cite{aniculaeseiVerification2016} checked robot-environment models with UPPAAL and synthesized safety monitors for the implementation.
Desai et al.~\cite{desaiCombining2017} combined model checking of core robot software in P with STL-based runtime monitoring of low-level control.
Schürmann et al.~\cite{schurmannEnsuring2017} verified the drivability of planned trajectories by combining reachability analysis with conformance testing.
Nigam and Talcott~\cite{nigamAutomating2023} automated recoverability proofs for runtime-assurance architectures of cyber-physical systems.
König et al.~\cite{konigSafe2024} applied model checking to an industrial behavior planner and validated counterexamples in simulation; their work targets the planning layer, whereas ours targets the trajectory-following control layer and its continuous dynamics.

\emph{Verification of Autoware.}
Miura et al.~\cite{miuraAutoware2019} provided the Autoware Toolbox with Simulink models of \Impl{Autoware} modules for model-based analysis and co-simulation.
Tran et al.~\cite{tranSafety2025} conducted runtime verification of \Impl{Autoware} using standardized critical scenarios but did not inspect the internals of the controller node.
To the best of our knowledge, ours is the first work to formally model (as a hybrid system) and monitor the trajectory-following controller of \Impl{Autoware}.

\section{Conclusion}

\Mod{We presented a model for the trajectory-following controller of \Impl{Autoware} and results of a hybrid verification approach.}
We adapted the $\GR$ model~\cite{bohrerFormal2019} to abstract the \Impl{Autoware} pure-pursuit lateral and PID longitudinal controllers as a hybrid program in dL.
We also developed a runtime monitoring method that instruments the \Impl{Autoware} controller node, orchestrates randomized simulations, and performs an offline evaluation of the constraints on the vehicle state that ensure the conformance to the $\GR$ model against the collected data.

The experimental results using two simulators of the \Impl{Autoware} toolchain answered the three research questions as follows:
\textbf{(RQ1)}~We confirmed the soundness of the monitoring approach.
\textbf{(RQ2)}~We characterized the effective trajectory-following accuracy of the \Impl{Autoware} implementation with the accuracy parameters $\Eps = 0.2$--$0.4$\,(m) and $\Eps_v = 0.8$\,(m/s).
\textbf{(RQ3)}~The method was also applicable to the more realistic \Impl{AWSIM} simulator, although the degree of conformance to the model was lower.

Future work includes 
investigating the root causes of accuracy drops and task failures,
\Mod{verifying other properties e.g. collision avoidance, and}
extending the approach to 
other practical controllers e.g. MPC-based ones.

\subsubsection*{\ackname}
The author thanks Pengfei Lin and Supakit Sroynam for their collaboration on the preliminary experiments, and Rose Bohrer and Stefan Mitsch for their help with \Impl{KeYmaera~X}.
This work was supported by JST, CREST Grant Number JPMJCR23M1, and JSPS, KAKENHI Grant Number 26K14771.

\bibliographystyle{splncs04}
\bibliography{AW-Mplex}

\begin{thebibliography}{10}
\providecommand{\url}[1]{\texttt{#1}}
\providecommand{\urlprefix}{URL }
\providecommand{\doi}[1]{https://doi.org/#1}

\bibitem{aniculaeseiVerification2016}
Aniculaesei, A., Arnsberger, D., Howar, F., Rausch, A.: Towards the {{Verification}} of {{Safety}}-critical {{Autonomous Systems}} in {{Dynamic Environments}}. Electronic Proceedings in Theoretical Computer Science  \textbf{232},  79--90 (Dec 2016). \doi{10.4204/EPTCS.232.10}

\bibitem{bohrerFormal2019}
Bohrer, R., Tan, Y.K., Mitsch, S., Sogokon, A., Platzer, A.: A {{Formal Safety Net}} for {{Waypoint-Following}} in {{Ground Robots}}. IEEE Robotics and Automation Letters  \textbf{4}(3),  2910--2917 (Jul 2019). \doi{10.1109/LRA.2019.2923099}

\bibitem{desaiCombining2017}
Desai, A., Dreossi, T., Seshia, S.A.: Combining {{Model Checking}} and {{Runtime Verification}} for {{Safe Robotics}}. In: {{RV}}. pp. 172--189. \LNCS{10548}, Springer (2017). \doi{10.1007/978-3-319-67531-2_11}

\bibitem{ferrandoROSMonitoring2020a}
Ferrando, A., Cardoso, R.C., Fisher, M., Ancona, D., Franceschini, L., Mascardi, V.: {{ROSMonitoring}}: {{A Runtime Verification Framework}} for {{ROS}}. In: Towards {{Autonomous Robotic Systems}}. pp. 387--399. \LNCS{12228}, Springer (2020). \doi{10.1007/978-3-030-63486-5_40}

\bibitem{fisherVerifying2013}
Fisher, M., Dennis, L., Webster, M.: Verifying autonomous systems. Communications of the ACM  \textbf{56}(9),  84--93 (Sep 2013). \doi{10.1145/2494558}

\bibitem{fritzschExperiences2021}
Fritzsch, J., Schmid, T., Wagner, S.: Experiences from {{Large-Scale Model Checking}}: {{Verifying}} a {{Vehicle Control System}} with {{NuSMV}}. In: {{ICST}}. pp. 372--382. IEEE, Porto de Galinhas, Brazil (2021). \doi{10.1109/ICST49551.2021.00049}

\bibitem{fultonKeYmaera2015}
Fulton, N., Mitsch, S., Quesel, J.D., V{\"o}lp, M., Platzer, A.: {{KeYmaera X}}: {{An Axiomatic Tactical Theorem Prover}} for {{Hybrid Systems}}. In: Felty, A.P., Middeldorp, A. (eds.) {{CADE-25}}. pp. 527--538. \LNCS{9195}, Springer (2015). \doi{10.1007/978-3-319-21401-6_36}

\bibitem{huangROSRV2014}
Huang, J., Erdogan, C., Zhang, Y., Moore, B., Luo, Q., Sundaresan, A., Rosu, G.: {{ROSRV}}: {{Runtime}} verification for robots. In: {{RV}}. pp. 247--254. \LNCS{8734}, Springer (2014). \doi{10.1007/978-3-319-11}

\bibitem{kalraDriving2016}
Kalra, N., Paddock, S.M.: Driving to safety: {{How}} many miles of driving would it take to demonstrate autonomous vehicle reliability? Transportation Research Part A: Policy and Practice  \textbf{94},  182--193 (Dec 2016). \doi{10.1016/j.tra.2016.09.010}

\bibitem{katoAutoware2018}
Kato, S., Tokunaga, S., Maruyama, Y., Maeda, S., Hirabayashi, M., Kitsukawa, Y., Monrroy, A., Ando, T., Fujii, Y., Azumi, T.: Autoware on {{Board}}: {{Enabling Autonomous Vehicles}} with {{Embedded Systems}}. In: {{ICCPS}}. pp. 287--296. IEEE (2018). \doi{10.1109/ICCPS.2018.00035}

\bibitem{konigSafe2024}
K{\"o}nig, L., Heinzemann, C., Griggio, A., Klauck, M., Cimatti, A., Henze, F., Tonetta, S., K{\"u}perkoch, S., Fassbender, D., Hanselmann, M.: Towards {{Safe Autonomous Driving}}: {{Model Checking}} a {{Behavior Planner}} during {{Development}}. In: {{TACAS}}. pp. 44--65. \LNCS{14571}, Springer (2024). \doi{10.1007/978-3-031-57249-4_3}

\bibitem{linReachFlow2020}
Lin, Q., Chen, X., Khurana, A., Dolan, J.M.: {{ReachFlow}}: {{An Online Safety Assurance Framework}} for {{Waypoint-Following}} of {{Self}}-driving {{Cars}}. In: {{IROS}}. pp. 6627--6632. IEEE (2020). \doi{10.1109/IROS45743.2020.9341122}

\bibitem{linSafe2022}
Lin, Q., Mitsch, S., Platzer, A., Dolan, J.M.: Safe and {{Resilient Practical Waypoint-Following}} for {{Autonomous Vehicles}}. IEEE Control Systems Letters  \textbf{6},  1574--1579 (2022). \doi{10.1109/LCSYS.2021.3125717}

\bibitem{loosAdaptive2011}
Loos, S.M., Platzer, A., Nistor, L.: Adaptive {{Cruise Control}}: {{Hybrid}}, {{Distributed}}, and {{Now Formally Verified}}. In: {{FM}}. pp. 42--56. \LNCS{6664}, Springer (2011). \doi{10.1007/978-3-642-21437-0_6}

\bibitem{luckcuckFormal2020}
Luckcuck, M., Farrell, M., Dennis, L.A., Dixon, C., Fisher, M.: Formal {{Specification}} and {{Verification}} of {{Autonomous Robotic Systems}}: {{A Survey}}. ACM Computing Surveys  \textbf{52}(5),  1--41 (Sep 2020). \doi{10.1145/3342355}

\bibitem{mehdipourFormal2023a}
Mehdipour, N., Althoff, M., Tebbens, R.D., Belta, C.: Formal methods to comply with rules of the road in autonomous driving: {{State}} of the art and grand challenges. Automatica  \textbf{152},  110692 (Jun 2023). \doi{10.1016/j.automatica.2022.110692}

\bibitem{mitschFormal2017}
Mitsch, S., Ghorbal, K., Vogelbacher, D., Platzer, A.: Formal verification of obstacle avoidance and navigation of ground robots. The International Journal of Robotics Research  \textbf{36}(12),  1312--1340 (Oct 2017). \doi{10.1177/0278364917733549}

\bibitem{mitschModelPlex2016}
Mitsch, S., Platzer, A.: {{ModelPlex}}: Verified runtime validation of verified cyber-physical system models. Formal Methods in System Design  \textbf{49}(1-2),  33--74 (Oct 2016). \doi{10.1007/s10703-016-0241-z}

\bibitem{miuraAutoware2019}
Miura, K., Tokunaga, S., Ota, N., Tange, Y., Azumi, T.: Autoware {{Toolbox}}: {{MATLAB}}/{{Simulink Benchmark Suite}} for {{ROS}}-based {{Self}}-driving {{Software Platform}}. In: International {{Workshop}} on {{Rapid System Prototyping}} ({{RSP}}). pp. 8--14. ACM (2019). \doi{10.1145/3339985.3358494}

\bibitem{nenchevCodeLevel2025}
Nenchev, V., Imrie, C., Gerasimou, S., Calinescu, R.: Code-{{Level Safety Verification}} for {{Automated Driving}}: {{A Case Study}}. In: {{FM}}. pp. 356--372. \LNCS{14934}, Springer (2025). \doi{10.1007/978-3-031-71177-0_22}

\bibitem{nigamAutomating2023}
Nigam, V., Talcott, C.: Automating {{Recoverability Proofs}} for {{Cyber-Physical Systems}} with {{Runtime Assurance Architectures}}. In: David, C., Sun, M. (eds.) Theoretical {{Aspects}} of {{Software Engineering}}. pp. 1--19. \LNCS{13931}, Springer (2023). \doi{10.1007/978-3-031-35257-7_1}

\bibitem{nilssonCorrectbyConstruction2016}
Nilsson, P., Hussien, O., Balkan, A., Chen, Y., Ames, A.D., Grizzle, J.W., Ozay, N., Peng, H., Tabuada, P.: Correct-by-{{Construction Adaptive Cruise Control}}: {{Two Approaches}}. IEEE Transactions on Control Systems Technology  \textbf{24}(4),  1294--1307 (Jul 2016). \doi{10.1109/TCST.2015.2501351}

\bibitem{othmanPublic2021}
Othman, K.: Public acceptance and perception of autonomous vehicles: A comprehensive review. AI and Ethics  \textbf{1}(3),  355--387 (Aug 2021). \doi{10.1007/s43681-021-00041-8}

\bibitem{platzerLogical2018}
Platzer, A.: Logical {{Foundations}} of {{Cyber-Physical Systems}}. Springer (2018). \doi{10.1007/978-3-319-63588-0}

\bibitem{prebetVerification2025}
Prebet, E., Teuber, S., Platzer, A.: Verification of {{Autonomous Neural Car Control}} with {{KeYmaera X}}. In: {{ABZ}}. pp. 288--307. \LNCS{15728}, Springer (Apr 2025). \doi{10.1007/978-3-031-94533-5_17}

\bibitem{rizaldiFormalising2017}
Rizaldi, A., Keinholz, J., Huber, M., Feldle, J., Immler, F., Althoff, M., Hilgendorf, E., Nipkow, T.: Formalising and {{Monitoring Traffic Rules}} for {{Autonomous Vehicles}} in {{Isabelle}}/{{HOL}}. In: {{IFM}}. pp. 50--66. \LNCS{10510}, Springer, Cham (2017). \doi{10.1007/978-3-319-66845-1_4}

\bibitem{schurmannEnsuring2017}
Schurmann, B., Hes, D., Eilbrecht, J., Stursberg, O., Koster, F., Althoff, M.: Ensuring drivability of planned motions using formal methods. In: {{ITSC}}. pp.~1--8. IEEE (2017). \doi{10.1109/ITSC.2017.8317647}

\bibitem{tangCollision2021}
Tang, Y., Zhou, Y., Liu, Y., Sun, J., Wang, G.: Collision {{Avoidance Testing}} for {{Autonomous Driving Systems}} on {{Complete Maps}}. In: 2021 {{IEEE Intelligent Vehicles Symposium}} ({{IV}}). pp. 179--185. IEEE, Nagoya, Japan (Jul 2021). \doi{10.1109/IV48863.2021.9575536}

\bibitem{tranSafety2025}
Tran, D.D., Tomita, T., Aoki, T.: Safety {{Analysis}} of {{Autonomous Driving Systems}}: {{A Simulation-Based Runtime Verification Approach}}. IEEE Transactions on Reliability pp. 1--15 (2025). \doi{10.1109/TR.2025.3561455}

\bibitem{tulcanHybrid2026}
Tulcan, R.F., Bohrer, R., Montacute, Y., Zhou, K., Kawamoto, Y., Hasuo, I.: Hybrid {{Spatiotemporal Logic}} for {{Automotive Applications}}: {{Modeling}} and {{Model-Checking}}. In: {{FM}}. pp. 337--356. \LNCS{16556}, Springer (2026). \doi{10.1007/978-3-032-26204-2_18}

\bibitem{viswanadhaAddressing2021}
Viswanadha, K., Indaheng, F., Wong, J., Kim, E., Kalvan, E., Pant, Y., Fremont, D.J., Seshia, S.A.: Addressing the {{IEEE AV Test Challenge}} with {{Scenic}} and {{VerifAI}}. In: {{AITest}}. pp. 136--142. IEEE (2021). \doi{10.1109/AITEST52744.2021.00034}

\bibitem{zhouSpecificationbased2023}
Zhou, Y., Sun, Y., Tang, Y., Chen, Y., Sun, J., Poskitt, C.M., Liu, Y., Yang, Z.: Specification-based {{Autonomous Driving System Testing}}. IEEE Transactions on Software Engineering pp. 1--19 (2023). \doi{10.1109/TSE.2023.3254142}

\end{thebibliography}

\appendix

\section{Proof of Proposition~\ref{prop}}
\label{a:proof}

\Mod{The \emph{simulation relation} follows the standard definition used in formal methods: Every state reachable by $\TF$, under the coordinate transformation, is also reachable by $\GR$.
The correspondence of initial states of $\TF$ and $\GR$ holds, since $\mathit{Init}$ is stated directly over the local variables $(x,y,v)$ that represent the coordinate-transformed state of $\TF$.}
Since the programs $\TF$ and $\GR$ evolve with either a discrete control step or a continuous dynamic phase, we verify that the state correspondence is preserved in each case.

For discrete steps, the correspondence follows from the assumptions.
$\mathsf{ctrlTF}$ updates the input $(\delta,a)$, which translates equivalently to the input $(k,a) := (\tfrac{\tan(\delta)}{\mathit{WB}},a)$ in $\GR$.
\Mod{By hypothesis, the translated $(k,a)$ satisfies $\mathit{Feas}$ and $\mathit{Go}$. Since $\mathsf{ctrlGR}$ allows any transition satisfying these constraints, the discrete-step correspondence holds.}

For continuous phases, we confirm that the dynamics equation of $\TF$ is equivalent to that of $\GR$ after the coordinate transformation.
We consider the case $v=1$; the general case is handled afterwards.
By the coordinate transformation, we have
\begin{equation*}
  \begin{bmatrix} x \\ y \end{bmatrix} = 
  \begin{bmatrix}
    \cos(\psi) (X_\mathrm{w}-X) + \sin(\psi) (Y_\mathrm{w}-Y) \\
    -\sin(\psi) (X_\mathrm{w}-X) + \cos(\psi) (Y_\mathrm{w}-Y) 
  \end{bmatrix}.
\end{equation*}
Differentiating both sides yields
\begin{equation*}
  \begin{bmatrix} x' \\ y' \end{bmatrix} = 
  \begin{bmatrix}
    \hspace{-7em} (-\sin(\psi) \psi' (X_\mathrm{w}-X) + \cos(\psi) \psi' (Y_\mathrm{w}-Y)) + \\
    \hspace{15em} (\cos(\psi) (-X') + \sin(\psi) (-Y')) \\
    \hspace{-7em} (-\cos(\psi) \psi' (X_\mathrm{w}-X) - \sin(\psi) \psi' (Y_\mathrm{w}-Y)) + \\
    \hspace{15em} (-\sin(\psi) (-X') + \cos(\psi) (-Y')) 
  \end{bmatrix}.
\end{equation*}
Using $\psi' = k$, $X' = \cos(\psi)$, $Y' = \sin(\psi)$, and trigonometric laws, it simplifies to
\begin{equation*}
  \begin{bmatrix} x' \\ y' \end{bmatrix} = 
  \begin{bmatrix}
    (ky) + 
    (-1) \\
    (-kx) + 
    (0) 
  \end{bmatrix},
\end{equation*}
which is equivalent to the dynamics equation specified in $\mathsf{plantGR}$ for $v=1$.
\Mod{Since $X'$, $Y'$, and $\psi'$ are each linear in $v$, so are the derived $x'$ and $y'$; the case for a general $v$ therefore follows by scaling the above equation by $v$. This gives $x' = v(ky-1)$ and $y' = -vkx$ specified in $\mathsf{plantGR}$.}

\Mod{Since the arguments for the discrete and continuous cases hold uniformly at each iteration, and the target waypoint of $\GR$ is fixed throughout, the correspondence of the whole execution holds by induction over the number of loop iterations, with the initial-state correspondence above as the base case.}

\section{Constraint Specifications in the $\GR$ Model}
\label{a:constraint}

The constraints are mostly the same as in the original model~\cite{bohrerFormal2019} but differ in a few ways. 
First, since we reduced the waypoints to a single one, its constraint is written only in the initial condition $\Init$. 
Also, whereas the original model constrains speed and acceleration in both the longitudinal and lateral directions, we simplified this to the longitudinal direction only.

The constraint $\mathit{Feas}$ in $\mathsf{ctrlGR}$ consists of two parts:
$$
  \mathit{Feas} ~~\equiv~~ \mathit{Ann} ~\land~ \mathit{Speed}
$$
Constraint $\mathit{Ann}$ states that the vehicle and the waypoint are within the $\Eps$-annulus.
This constraint represents an annulus of width $\Eps$ around the arc of curvature $k$ connecting the ego vehicle at $(0,0)$ and the waypoint at $(x,y)$, which is defined by $k = \tfrac{2y}{x^2+y^2}$ as in pure-pursuit control; the annulus is obtained by adding multiple $\Eps$ terms into this arc constraint.
$$
  \mathit{Ann} ~~\equiv~~
  |k| \Eps \leq 1 
  ~\land~ 
  \left|\frac{k (x^2 + y^2 - \Eps^2)}{2} - y\right| < \Eps
$$
%
Constraint $\mathit{Speed}$ states that the speed target bounds $v_l$ and $v_h$ are non-negative; it also states their ordering, and the relationship between the interval width $v_h-v_l$ and the speed change over the step duration $T$.
The target speed is non-negative, and acceleration/deceleration is feasible within the step duration.
\begin{align*}
  \mathit{Speed} ~~\equiv~~&
  0 \leq v_l < v_h ~\land~ A T \leq v_h - v_l ~\land~ B T \leq v_h - v_l
\end{align*}

Basic constraints on the velocity and acceleration are prepared as $\mathit{GoA}$ and $\mathit{GoB}$.
The conjunction of the two constraints requires that either the speed is already within the target range, or there is sufficient longitudinal distance to accelerate (with $A$) or decelerate (with $-B$) to the target within the step duration $T$.
\begin{align*}
  \mathit{GoA} ~~\equiv~~&
  v_l \leq v ~\lor~ 
  (1 + |k|\Eps)^2 \left(v T + \frac{a}{2} T^2 + \frac{v_l^2 - (v + a T)^2}{2 A}\right) + \Eps \leq x \\
  \mathit{GoB} ~~\equiv~~&
  v \leq v_h ~\lor~ 
  (1 + |k|\Eps)^2 \left(v T + \frac{a}{2} T^2 + \frac{(v + a T)^2 - v_h^2}{2 B}\right) + \Eps \leq x
\end{align*}

The constraint $\mathit{Go}$ in $\mathsf{ctrlGR}$ specifies the requirement for controlled accelerations $a$.
$\mathit{GoA}'$/$\mathit{GoB}'$ is a modification of $\mathit{GoA}$/$\mathit{GoB}$ to bound the velocity $v+aT$;
the last term is modified for ease of proving.
$\mathit{Go}$ also appends basic constraints on $a$.
\begin{align*}
  \mathit{GoA}' ~~\equiv~~&
  (v_l \leq v ~\land~ v_l \leq v + a T)
  ~\lor~ \\
  & (1 + |k|\Eps)^2 \left(A (2 v T + a T^2) + (v_l^2 - (v+aT)^2)\right) \leq 2 A (x-\Eps) \\
  \mathit{GoB}' ~~\equiv~~&
  (v \leq v_h ~\land~ v + a T \leq v_h)
  ~\lor~ \\
  & (1 + |k|\Eps)^2 \left(B (2 v T + a T^2) + ((v+aT)^2 - v_h^2)\right) \leq 2 B (x-\Eps) \\
  %
  \mathit{Go} ~~\equiv~~&
  -B \leq a \leq A ~\land~ v + a T \geq 0 ~\land~ 
  \mathit{GoA}' ~\land~ \mathit{GoB}'
\end{align*}

Based on the above components, the following constraints are used in the formulas stating safety and liveness properties.
\begin{align*}
  \mathit{Inv} ~~\equiv~~&
  v \geq 0 ~\land~ 
  \mathit{Feas} ~\land~ \mathit{GoA} ~\land~ \mathit{GoB} \\
  \mathit{Init} ~~\equiv~~&
  T>0 ~\land~ \Eps>0 ~\land~ A>0 ~\land~ B>0 ~\land~ \mathit{Inv}
\end{align*}

\section{Artefact}
\label{a:artefact}

\subsection{Badge Claims}

We claim two badges:
(i)~\textbf{Artefacts Available} and
(ii)~\textbf{Artefacts Functional}.

\vspace{.5em}
\noindent
\emph{Availability}.
We archived the artefact package at: \url{https://doi.org/10.5281/zenodo.21510141}

\vspace{.5em}
\noindent
\emph{Functional outcomes}.
We consider the following functional outcomes: \\ 

\noindent
$\mathsf{F}_1$~The proofs of the safety and liveness properties of the $\GR$ model (described at the beginning of Sect.~\ref{s:verif}) are validated using \Impl{KeYmaera~X}.

\vspace{1em}
\noindent
$\mathsf{F}_2$~The \Impl{scenario\_simulator\_v2} tool in the \Impl{Autoware} toolchain generates the log files for offline monitoring as described in Sect.~\ref{s:verif:monitor}.
This consists of two sub-outcomes, depending on the backend used by \Impl{scenario\_simulator\_v2}.
\begin{description}
  \item[$\mathsf{F}_{2.1}$] Simulation with the \Impl{Simple} backend.
  \item[$\mathsf{F}_{2.2}$ (optional)] Simulation with the \Impl{AWSIM} tool based on the Unity physics engine.
  It depends heavily on the host PC environment. Although we have taken the time to prepare a Docker image and instructions for using it, we leave verification of $\mathsf{F}_{2.2}$ optional for reviewers.
\end{description}

\vspace{1em}
\noindent
$\mathsf{F}_3$~The offline monitoring of the simulation log data described in Sect.~\ref{s:verif:monitor}.

\subsection{Quick Start}

This section describes the contents of the artefact package and the host environment required to reproduce the functional outcomes.
As a sanity check, we describe how to pull and run some Docker containers (for $\mathsf{F}_{2.1}$ and $\mathsf{F}_3$); 
verification of the other containers is omitted here, as it involves more effort.
The concrete steps for each outcome are given in Sect.~\ref{s:functional-evaluation}.

\subsubsection{Artefact Contents}

\begin{figure}[!h]
\centering
\begin{mdframed}
\renewcommand*\DTstylecomment{\rmfamily}
\dirtree{%
.1 artefacts/.
.2 article.pdf \DTcomment{This document (submitted version)}\\
.
.2 GroundRobotX/ \DTcomment{The $\GR$ model specification in \Impl{KeYmaera~X}}.
.3 Makefile.
.3 *.kyx.txt \DTcomment{Templated \Impl{KeYmaera~X} file}.
.3 *.kyx \DTcomment{\Impl{KeYmaera~X} file}.
.3 *.kyx.kyp \DTcomment{Proof log}.
.3 docker/.
.4 Dockerfile \DTcomment{Dockerfile for the first image}.
.4 entrypoint.sh \DTcomment{Entrypoint script}.
.4 run.sh \DTcomment{Running script}\\
.
.2 autoware/ \DTcomment{Meta repository of the \Impl{Autoware} toolchain}.
.3 src/ \DTcomment{Source files of \Impl{Autoware} and related tools}.
.4 universe/ \DTcomment{Source files of (modified) \Impl{Autoware Universe}}.
.4 simulator/ \DTcomment{Source files of (modified) \Impl{scenario\_simulator\_v2}}.
.3 docker/ \DTcomment{Parent image definitions of the second Docker image}.
.3 docker\_vnc/ \DTcomment{Definition of the VNC layer of the second image} \\
.
.2 run\_simulator.sh \DTcomment{Running script for the second and third images}\\
.
.2 autoware\_control\_monitor/ \DTcomment{Offline monitoring script}.
.3 main.py \DTcomment{Main Python script}.
.3 docker/.
.4 Dockerfile \DTcomment{Dockerfile for the fourth image}.
.4 entrypoint.sh \DTcomment{Entrypoint script}.
.4 run.sh \DTcomment{Running script}\\
.
.2 simple100/ \DTcomment{Log files of the simulation with \Impl{Simple}}.
.3 $i\,$/ \DTcomment{Subdirectory corresponding to the $i$-th scenario}.
.4 context.yaml \DTcomment{Summary}.
.4 *.csv \DTcomment{Log files}.
.4 sample\_awsim\_mod.xosc \DTcomment{\Impl{OpenSCENARIO} file}.
.3 stat\_simple.xlsx \DTcomment{Excel file summarizing the statistics}.
.2 awsim100/ \DTcomment{Log files of the simulation with \Impl{AWSIM}}.
.3 $i\,$/ \DTcomment{Subdirectory corresponding to the $i$-th scenario}.
.3 stat\_awsim.xlsx.
}
\end{mdframed}
\caption{Directory structure of the artefact package.}
\label{f:dirtree}
\end{figure}

The package contains the source code of the artefacts and the experimental data.
The directory structure of the artefact package is shown in Fig.~\ref{f:dirtree}.

For functional verification, we have prepared four Docker images.
\begin{enumerate}
\item \url{ghcr.io/cpcp-lab/artefacts:kyx-sefm2026} for $\mathsf{F}_{1}$ (\verb|amd64|; about 750\,MB).
\item \url{ghcr.io/cpcp-lab/aw-tools:scenario-simulator-sefm2026} for $\mathsf{F}_{2.1}$\\ (\verb|amd64|/\verb|arm64|; about 10\,GB).
\item \url{ghcr.io/cpcp-lab/aw-tools:scenario-simulator-cuda-sefm2026} for $\mathsf{F}_{2.2}$ (optional; \verb|amd64|; about 35\,GB).
\item \url{ghcr.io/cpcp-lab/artefacts:monitor-sefm2026} for $\mathsf{F}_{3}$ (\verb|amd64|/\verb|arm64|;\\ about 350\,MB).
\end{enumerate}
The scripts included in the package pull and run these images.

\subsubsection{Host Environment}

All four Docker images assume an interactive shell environment with the \verb|docker| command available, but they differ in supported architecture, GPU requirements, GUI display method, and additional license activation steps. 

The usage conditions of each image are summarized below.
\begin{enumerate}
\item \texttt{ghcr.io/cpcp-lab/artefacts:kyx-sefm2026} ($\mathsf{F}_{1}$):
  \begin{itemize}
    \item It is launched with \texttt{--platform=linux/amd64}, so on ARM hosts (e.g. Apple Silicon)
      it runs via emulation.
    \item When the container is invoked, Wolfram Engine is installed automatically (which will take some time); if it is not yet activated, an interactive activation using a Wolfram ID is required.
    See \url{https://support.wolfram.com/46072} for details.
  \end{itemize}

\item \texttt{ghcr.io/cpcp-lab/aw-tools:scenario-simulator-sefm2026} ($\mathsf{F}_{2.1}$):
  \begin{itemize}
    \item An \texttt{arm64} build is available in addition to \texttt{amd64}, and operation on macOS
      (Apple Silicon) has been confirmed.
    \item X11 forwarding does not work reliably on macOS, so the VNC mode with the \texttt{--vnc} option is recommended there.
  \end{itemize}

\item \texttt{ghcr.io/cpcp-lab/aw-tools:scenario-simulator-cuda-sefm2026} ($\mathsf{F}_{2.2}$):
  \begin{itemize}
    \item It requires a Linux host equipped with at least 32\,GB RAM, an NVIDIA GPU with at least 12\,GB VRAM, and the corresponding NVIDIA driver.
    \item Since the GPU must be exposed to the container, CUDA and the NVIDIA Container Toolkit
      must be installed on the host.
    We do not describe these installation procedures in this document.
  \end{itemize}

\item \texttt{ghcr.io/cpcp-lab/artefacts:monitor-sefm2026} ($\mathsf{F}_{3}$):
  \begin{itemize}
    \item An \texttt{arm64} build is available in addition to \texttt{amd64}, and operation on macOS
      (Apple Silicon) has been confirmed.
  \end{itemize}
\end{enumerate}

We verified the artefact on the following three host machines.
The third image was not tested on M3.
\begin{description}
\item[M1] CPU: AMD Ryzen 9 7950X (16 cores/32 threads, 4.5--5.7\,GHz); RAM: 64\,GB; GPU: NVIDIA GeForce RTX 5060 Ti; OS: Ubuntu 22.04; Docker~29.4.1; CUDA 12.9.
\item[M2] CPU: AMD Ryzen 9 5900XT (12 cores/24 threads, 3.7--4.8\,GHz); RAM: 64\,GB; GPU: NVIDIA GeForce RTX 4070; OS: Ubuntu 22.04; Docker 29.5.2; CUDA 13.0.
\item[M3] MacBook Pro (2023); CPU: Apple Silicon M2 (8 cores: 4 performance + 4 efficiency, up to 3.49\,GHz); RAM: 32\,GB; OS: macOS 15.7.7; Docker Desktop~4.52.0.
\end{description}

\subsubsection{Sanity Check}

We denote by \verb|${ARTEFACTS}| the directory where the artefact package is placed.
The running scripts for the Docker containers are expected to be executed from \verb|${ARTEFACTS}|.

We describe the steps to pull and run the Docker containers for $\mathsf{F}_{2.1}$ and $\mathsf{F}_3$.

\vspace{.5em}
\noindent
\emph{$\mathsf{F}_{2.1}$}.
The script \verb|run_simulator.sh| invokes the second Docker container as follows.
Depending on the host environment, users may run it in either X11 mode or VNC mode.

\begin{lstlisting}[style=shell]
(host)$ @cd ${ARTEFACTS}@
(host)$ @./run_simulator.sh --help@
Usage: run.sh [OPTIONS]
Options:
  --help/-h            Display this help message
  --vnc                Use VNC mode (default: X11 mode)
  --geometry           VNC desktop resolution WxH (default: 1920x1200)
  --display            VNC display number (default: 1, port: 5901)
  --workspace          Directory to mount into /workspace (default: current directory)
  --map-path           Mount map files into /autoware_map
  --data-path          Mount data files into /autoware_data or /autoware_out
  --cuda               Enable NVIDIA GPU support
  --image              Docker image to use (overrides built-in image names)
  --pull-latest-image  Pull the latest image before starting
\end{lstlisting}

As a sanity check, we simply pull and run the container in VNC mode.
The total image size is approximately 10\,GB, and thus requires a corresponding amount of time and free storage.
\begin{lstlisting}[style=shell]
(host)$ @./run_simulator.sh --vnc@
Launching Autoware VNC environment
Connect VNC client to: localhost:5901  (password: ubuntu)
IMAGE: ghcr.io/cpcp-lab/aw-tools:scenario-simulator-sefm2026
WORKSPACE PATH(mounted): ${ARTEFACTS}:/workspace
DATA PATH(mounted): ${ARTEFACTS}/data:/autoware_data
-----------------------------------------------------------------
+ '[' false = true ']'
++ cat /etc/timezone
++ echo UTC
+ docker run -it --rm -p 5901:5901 -e LOCAL_UID=1001 -e LOCAL_GID=1000 -e LOCAL_USER=user -e LOCAL_GROUP=group -e NVIDIA_DRIVER_CAPABILITIES=all -e TZ=UTC -v ${ARTEFACTS}:/workspace -v ${ARTEFACTS}/data:/autoware_data:rw ghcr.io/cpcp-lab/aw-tools:scenario-simulator-sefm2026 /bin/bash -c 'vncserver :1 -geometry 1920x1200 -depth 24 -localhost no; exec /bin/bash'
Unable to find image 'ghcr.io/cpcp-lab/aw-tools:scenario-simulator-sefm2026' locally
scenario-simulator-sefm2026: Pulling from cpcp-lab/aw-tools
7c7373a68ead: Pull complete
5dbd4e32690a: Pull complete
....
b9e5cf78b120: Pull complete
Digest: sha256:cfb08e977442aeef8beaca10b3ede88ed0286b004fc0837b794655335ab2871d
Status: Downloaded newer image for ghcr.io/cpcp-lab/aw-tools:scenario-simulator-sefm2026
Starting with user: user >> UID 1001, GID: 1000
....

New Xtigervnc server '037777777776:1 (user)' on port 5901 for display :1.
Use xtigervncviewer -SecurityTypes VncAuth,TLSVnc -passwd /home/user/.vnc/passwd 037777777776:1 to connect to the VNC server.
.... A VNC server has been started on port 5901.

user%@%037777777776:/autoware$ @exit@
\end{lstlisting}

\vspace{.5em}
\noindent
\emph{$\mathsf{F}_{3}$}.
The script \verb|autoware_control_monitor/docker/run.sh| is prepared.
Simply executing it pulls and runs the fourth container as follows.
If successful, a wrapper script prompting for a choice will be invoked.
\begin{lstlisting}[style=shell]
(host)$ @cd ${ARTEFACTS}@
(host)$ @./autoware_control_monitor/docker/run.sh --help@
Usage: run.sh [OPTIONS] [LAUNCH_CMD]
Options:
  --help/-h            Display this help message
  --workers N          Number of parallel workers for batch processing
.... You may consider setting --workers N if many CPU cores are available (default: N=4)

(host)$ @./autoware_control_monitor/docker/run.sh@
-----------------------------------------------------------------
IMAGE: ghcr.io/cpcp-lab/artefacts:monitor-sefm2026
WORKSPACE PATH(mounted): ${ARTEFACTS}:/artefacts
LAUNCH_CMD:
-----------------------------------------------------------------
+ docker run -it --rm --net=host -v ${ARTEFACTS}:/artefacts ghcr.io/cpcp-lab/artefacts:monitor-sefm2026
Unable to find image 'ghcr.io/cpcp-lab/artefacts:monitor-sefm2026' locally
monitor-sefm2026: Pulling from cpcp-lab/artefacts
7c7373a68ead: Pull complete
....
a3dddf233d94: Pull complete
Digest: sha256:7430e09fe4dd09282316d15c2903cea54264c21b52a0e0b71c9c4e78cf55d8fc
Status: Downloaded newer image for ghcr.io/cpcp-lab/artefacts:monitor-sefm2026
Please select an operation to perform:
1. Generate Fig. 5
2. Generate a plot figure with parameter specifications
3. Batch monitoring with parameter specifications (may take several minutes)
4. Run batch for all parameter combinations (may take more than an hour)
5. Command line
6. Exit
Choice [1-6]: @6@
\end{lstlisting}

\subsection{Functional Evaluation}
\label{s:functional-evaluation}

In this section, we describe the process of the functional evaluation on the Docker containers.

\subsubsection{$\mathsf{F}_1$: Proof Validation with \Impl{KeYmaera~X}}

\textbf{Time required: about 30--60 minutes.}

\vspace{.5em}
\noindent
The script \verb|GroundRobotX/docker/run.sh| invokes the first Docker container as follows.
Invoking this container automatically downloads and installs Wolfram Engine, which is not bundled in the Docker image itself.
Since this step depends on the distribution channel of Wolfram Research, it may fail if the channel becomes unavailable in the future.
On ARM hosts, it runs in emulation mode, which makes processing several times slower.
\begin{lstlisting}[style=shell]
(host)$ @cd ${ARTEFACTS}@
(host)$ @./GroundRobotX/docker/run.sh@
.... This will pull the Docker image and run it.
.... At first, it downloads and installs the WolframEngine package, and then the container shell will be invoked.
-----------------------------------------------------------------
IMAGE: ghcr.io/cpcp-lab/artefacts:kyx-sefm2026
WORKSPACE PATH(mounted): ${ARTEFACTS}:/artefacts
LAUNCH_CMD:
-----------------------------------------------------------------
+ docker run -it --rm --net=host --platform=linux/amd64 -v ${ARTEFACTS}:/artefacts ghcr.io/cpcp-lab/artefacts:kyx-sefm2026
Installing Wolfram Engine...
Reading package lists... Done
Building dependency tree... Done
Reading state information... Done
Note, selecting 'wolfram-engine' instead of '/artefacts/GroundRobotX/docker/WolframEngine/wolfram-engine.deb'
The following NEW packages will be installed:
  wolfram-engine
.... Here, it downloads and installs the WolframEngine package.
....
Wolfram Engine installation completed.
.... Then there will be an authentication interaction.
....
Please select an operation to perform:
1. Single proof validity check
2. Batch proof validation
3. Command line
4. Exit
.... Select a number to run a function.
Choice [1-4]: @1@
Please select the .kyx file to verify:
1. Invariance.kyx
2. LivenessEvt.kyx
3. LivenessHigh.kyx
4. Liveness.kyx
5. LivenessLow.kyx
6. LivenessMid.kyx
7. Safety.kyx
Choice [1-7]: @1@
make: Entering directory '/artefacts/GroundRobotX'
Proving Invariance.kyx...
PROVED GroundRobotX/Invariance: tactic=GroundRobotX/Invariance: Proof,tacticsize=130,budget=0[s],duration=11865[ms],qe=9514[ms],rcf=0,steps=21210
make: Leaving directory '/artefacts/GroundRobotX'

Please select an operation to perform:
1. Single proof validity check
2. Batch proof validation
3. Command line
4. Exit
.... This will validate all .kyx files, which takes several minutes
Choice [1-4]: @2@
make: Entering directory '/artefacts/GroundRobotX'
Proving Invariance.kyx...
PROVED GroundRobotX/Invariance: tactic=GroundRobotX/Invariance: Proof,tacticsize=130,budget=0[s],duration=6795[ms],qe=3979[ms],rcf=0,steps=5442
Proving LivenessEvt.kyx...
PROVED GroundRobotX/LivenessEvt: tactic=GroundRobotX/LivenessEvt: Proof,tacticsize=54,budget=0[s],duration=1372[ms],qe=188[ms],rcf=0,steps=506
Proving LivenessHigh.kyx...
PROVED GroundRobotX/LivenessHigh: tactic=GroundRobotX/LivenessHigh: Proof,tacticsize=53,budget=0[s],duration=63439[ms],qe=59867[ms],rcf=0,steps=23207
Proving Liveness.kyx...
PROVED GroundRobotX/Liveness: tactic=GroundRobotX/Liveness: Proof,tacticsize=9,budget=0[s],duration=786[ms],qe=0[ms],rcf=0,steps=203
Proving LivenessLow.kyx...
PROVED GroundRobotX/LivenessLow: tactic=GroundRobotX/LivenessLow: Proof,tacticsize=58,budget=0[s],duration=57807[ms],qe=54129[ms],rcf=0,steps=19301
Proving LivenessMid.kyx...
PROVED GroundRobotX/LivenessMid: tactic=GroundRobotX/LivenessMid: Proof,tacticsize=46,budget=0[s],duration=11289[ms],qe=9022[ms],rcf=0,steps=9918
Proving Safety.kyx...
PROVED GroundRobotX/Safety: tactic=GroundRobotX/Safety: Proof,tacticsize=18,budget=0[s],duration=2069[ms],qe=1443[ms],rcf=0,steps=761

make: Leaving directory '/artefacts/GroundRobotX'
Please select an operation to perform:
1. Single proof validity check
2. Batch proof validation
3. Command line
4. Exit
Choice [1-4]: @4@
\end{lstlisting}

The validity of \inlst{Invariance.kyx} and \inlst{Safety.kyx} implies that of the safety formula in Sect.~\ref{s:verif}.
The validity of \inlst{Liveness.kyx}, which depends on that of
\inlst{LivenessEvt.kyx}, \inlst{LivenessHigh.kyx}, \inlst{LivenessLow.kyx}, and \inlst{LivenessMid.kyx}, implies the validity of the liveness formula in Sect.~\ref{s:verif}.

\subsubsection{$\mathsf{F}_{2.1}$: Simulation with \Impl{Autoware}}

\textbf{Time required: about 10--30 minutes.}

\vspace{.5em}
\noindent
Depending on the host environment, users may run the script 
\lstinline|run_simulator.sh| in either X11 mode or VNC mode.

\begin{figure}[t]
  \centering
  \includegraphics[width=0.75\textwidth]{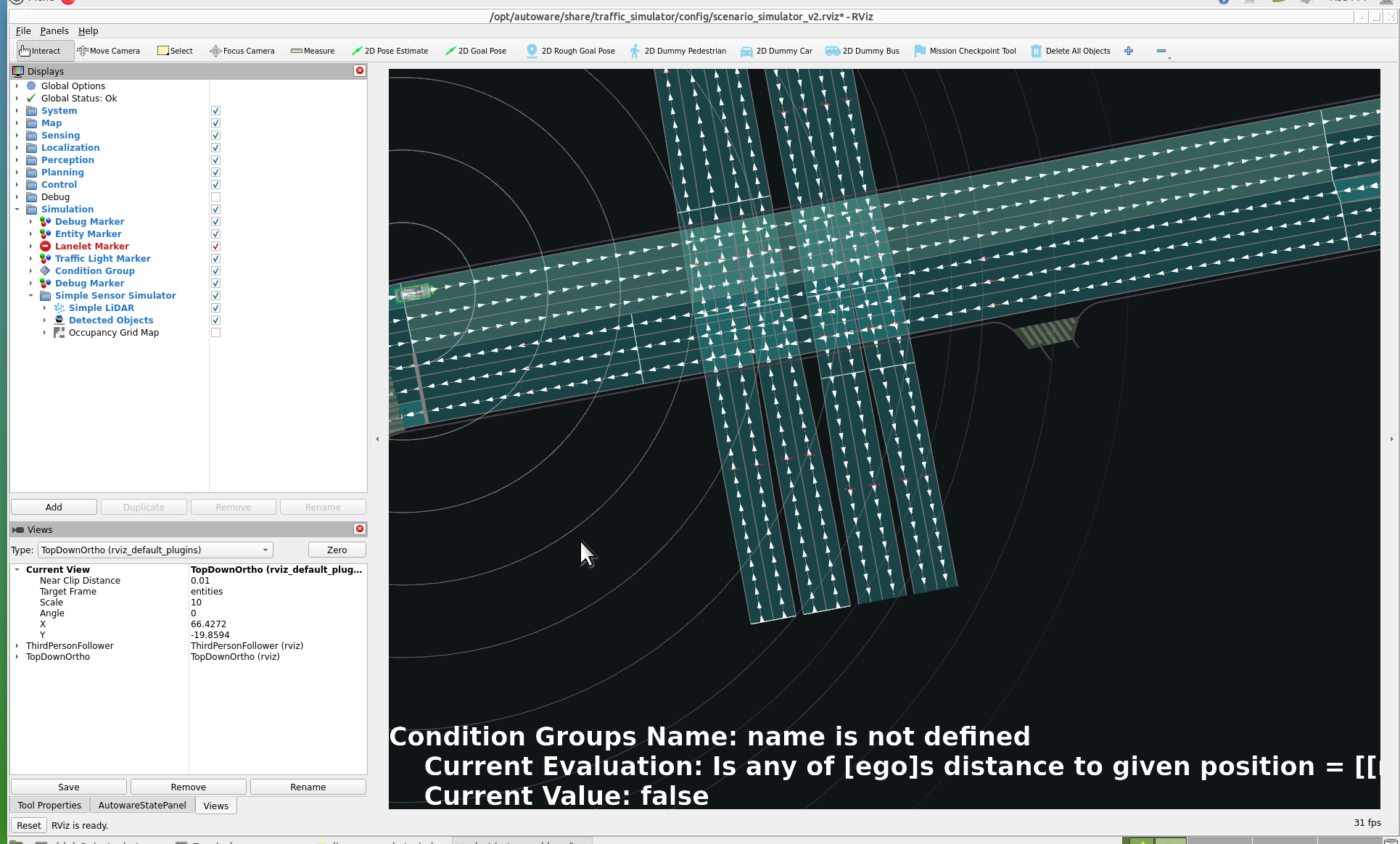}
  \caption{The \Impl{RViz} window of the simulator.}
  \label{f:rviz}
\end{figure}

\vspace{.5em}
\noindent
\emph{X11 Mode}.
This mode is recommended for Linux hosts.
\begin{lstlisting}[style=shell]
(host)$ @./run_simulator.sh /simulation.sh --s1@
.... This command will run three scenarios in sequence.
....
+ docker run -it --rm --net=host -e LOCAL_UID=1001 -e LOCAL_GID=1000 -e LOCAL_USER=user -e LOCAL_GROUP=group -e DISPLAY=:0 -v /tmp/.X11-unix/:/tmp/.X11-unix -e XAUTHORITY=/home/user/.Xauthority -e XDG_RUNTIME_DIR=/run/user/1001 -e NVIDIA_DRIVER_CAPABILITIES=all -e TZ=UTC -v ${ARTEFACTS}:/workspace -v ${ARTEFACTS}/data:/autoware_data:rw ghcr.io/cpcp-lab/aw-tools:scenario-simulator-sefm2026 /simulation.sh --s1
....
Running simple simulator...
ARCHITECTURE_TYPE: awf/universe/20240605
SENSOR_MODEL: sample_sensor_kit
VEHICLE_MODEL: sample_vehicle
....
[INFO] [launch]: Default logging verbosity is set to INFO
architecture_type                   := awf/universe/20240605
autoware_launch_file                := planning_simulator.launch.xml
....
[INFO] [test_runner.py-1]: process started with pid [66]
[INFO] [logger.py-2]: process started with pid [68]
....
\end{lstlisting}
If successful, the window shown in Fig.~\ref{f:rviz} will be opened and the movement of the vehicle will be shown.

As a result, a set of log files per scenario is output under the directory \lstinline|${ARTEFACTS}/data/test_runner-${TIMESTAMP}|.

The following command will run 163 scenarios in sequence. It will take several hours, so performing this verification is optional.
\begin{lstlisting}[style=shell]
(host)$ @./run_simulator.sh /simulation.sh --s100@
\end{lstlisting}

\vspace{.5em}
\noindent
\emph{VNC Mode}.
In case the X11 server connection does not work properly, we can invoke a VNC server in the container and connect to it from the host.
This mode is recommended for macOS hosts.
\begin{lstlisting}[style=shell]
(host)$ @cd ${ARTEFACTS}@
(host)$ @./run_simulator.sh --vnc@
....
New Xtigervnc server '037777777776:1 (user)' on port 5901 for display :1.
Use xtigervncviewer -SecurityTypes VncAuth,TLSVnc -passwd /home/user/.vnc/passwd 037777777776:1 to connect to the VNC server.
.... A VNC server has been started on port 5901.

user%@%037777777776:/autoware$
\end{lstlisting}
Users can connect from a host VNC viewer by accessing \url{vnc://localhost:5901}.
For example, on macOS, users can select the \verb|Go| $\to$ \verb|Connect to Server...| menu from the menu bar of \Impl{Finder}, enter the address in the dialog, and enter the password ``ubuntu''.
Then, the desktop window (\Impl{Screen Sharing} application) will open.
On the desktop, open the terminal (by clicking the icon of the Terminal application) and execute the simulation script.
\begin{lstlisting}[style=shell]
(vnc)$ @/simulation.sh --s1@
\end{lstlisting}
As with the X11 mode, the \Impl{RViz} window (Fig.~\ref{f:rviz}) will open, and log files are output under the directory \lstinline|${ARTEFACTS}/data/test_runner-${TIMESTAMP}|.

\subsubsection{$\mathsf{F}_{2.2}$: Simulation with \Impl{Autoware} using the \Impl{AWSIM} Backend}

\begin{figure}[t]
  \centering
  \includegraphics[width=0.45\textwidth]{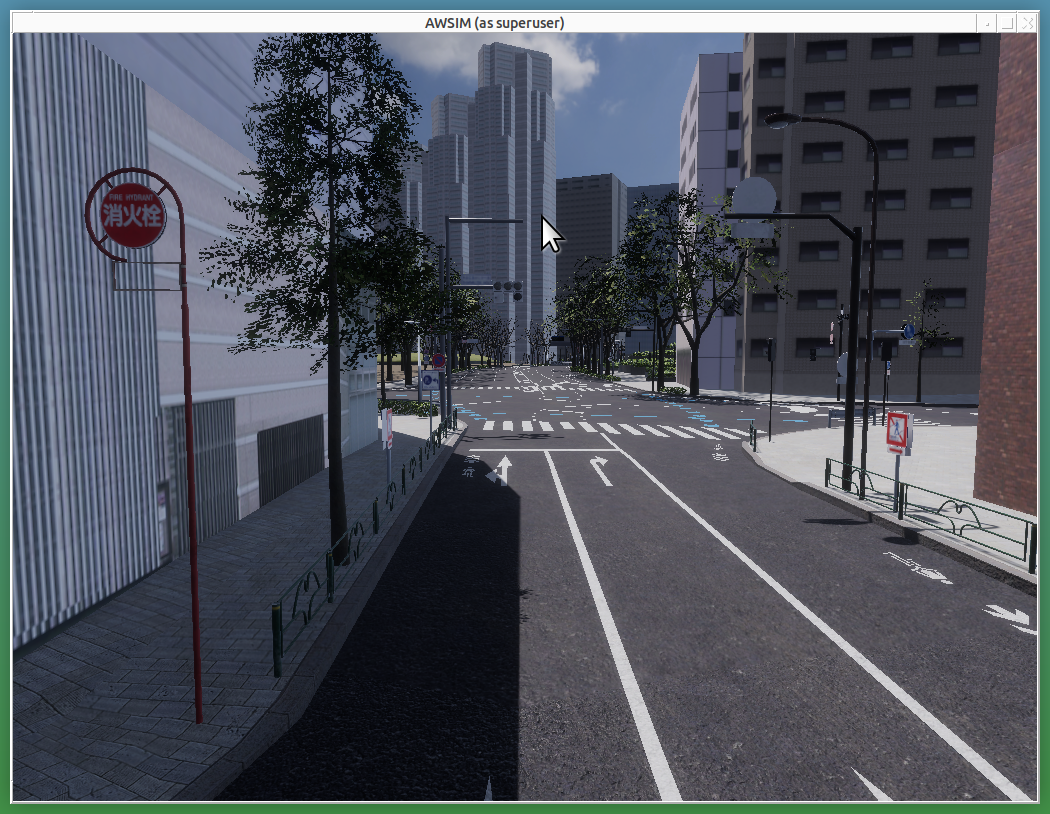}
  \hspace{.5em}
  \includegraphics[width=0.45\textwidth]{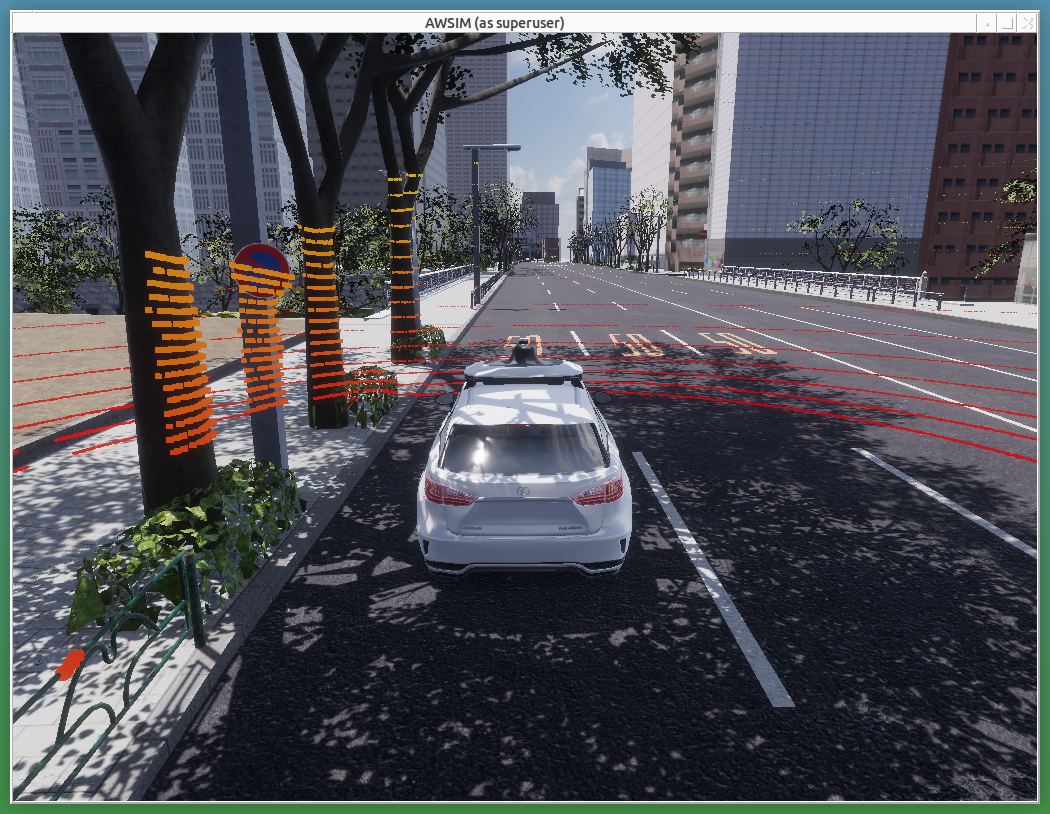}
  \caption{Screenshots of \Impl{AWSIM}.}
  \label{f:awsim}
\end{figure}

This outcome can be demonstrated in the third Docker container.
To do so, run the \\
\lstinline|run_simulator.sh| script with the \verb|--cuda| option, and
the \inlst{simulation.sh} script in the container with the \inlst{--awsim} option.
Otherwise, the usage is the same as for $\mathsf{F}_{2.1}$.
If successful, the simulation will run and the \Impl{AWSIM} window (Fig.~\ref{f:awsim}) will open.

\subsubsection{$\mathsf{F}_{3}$: Offline Monitoring}

\textbf{Time required: about 30--90 minutes.}

\vspace{.5em}
\noindent
The script \verb|autoware_control_monitor/docker/run.sh| invokes the fourth \AB Docker container as follows.

\begin{lstlisting}[style=shell]
(host)$ @cd ${ARTEFACTS}@
.... You may consider setting --workers N if many CPU cores are available (default: N=4)
(host)$ @./autoware_control_monitor/docker/run.sh --workers 12@
-----------------------------------------------------------------
IMAGE: ghcr.io/cpcp-lab/artefacts:monitor-sefm2026
WORKSPACE PATH(mounted): ${ARTEFACTS}:/artefacts
LAUNCH_CMD: --workers 12
-----------------------------------------------------------------
+ docker run -it --rm --net=host -v ${ARTEFACTS}:/artefacts ghcr.io/cpcp-lab/artefacts:monitor-sefm2026 --workers 12
Please select an operation to perform:
1. Generate Fig. 5
2. Generate a plot figure with parameter specifications
3. Batch monitoring with parameter specifications (may take several minutes)
4. Run batch for all parameter combinations (may take more than an hour)
5. Command line
6. Exit
.... Select a number to run a function.
Choice [1-6]: @1@
Please select the dataset:
1. Simple
2. AWSIM
Choice [1-2]: @1@
0([-1.500,0.842],[0.000,0.125]): !I18; Rok; (1&0&1&1)&1&1&1; !V2
1([-1.500,0.842],[0.000,0.125]): !I18; Rok; (1&0&1&1)&1&1&1; !V2
.... 
113: skipped (waypoint not reached or passed)
total(6 windows):   !Is: 2  !Rs: 0(0 unsound)   6/6* (106 skipped)
The plot has been saved to /artefacts/autoware_control_monitor/out.png
.... A plot is output as ${ARTEFACTS}/autoware_control_monitor/out.png

Please select an operation to perform:
1. Generate Fig. 5
2. Generate a plot figure with parameter specifications
3. Batch monitoring with parameter specifications (may take several minutes)
4. Run batch for all parameter combinations (may take more than an hour)
5. Command line
6. Exit
Choice [1-6]: @3@
Please select the dataset:
1. Simple
2. AWSIM
Choice [1-2]: @1@
Please select the accuracy setting:
1. eps=0.1, eps_v=0.4
2. eps=0.2, eps_v=0.8
3. eps=0.4, eps_v=1.6
Choice [1-3]: @2@
/artefacts/autoware_control_monitor/../simple100/20: skipped (marker_log.csv is empty)
.... All scenarios are processed with the given accuracy setting.
1:	!Is: 3	!Rs: 3(0 unsound)	9/12 (98 skipped)
2:	!Is: 1	!Rs: 1(0 unsound)	13/14 (2 skipped)
....
163:	!Is: 2	!Rs: 0(0 unsound)	21/21* (90 skipped)
total(157 runs): !Is: 249	!Rs: 151(0 unsound)	2227/2406 (7409 skipped)

Please select an operation to perform:
1. Generate Fig. 5
2. Generate a plot figure with parameter specifications
3. Batch monitoring with parameter specifications (may take several minutes)
4. Run batch for all parameter combinations (may take more than an hour)
5. Command line
6. Exit
Choice [1-6]: @4@
Please select the dataset:
1. Simple
2. AWSIM
Choice [1-2]: @1@
eps=0.0125 eps-v=0.05
eps=0.0125 eps-v=0.1
.... 
eps=1.6 eps-v=3.2
.... The result is output in the ${ARTEFACTS}/data/monitor_${TIMESTAMP} directory.
\end{lstlisting}
The result of the last batch process is used to produce the Excel files,\\ 
\lstinline|stat_simple.xlsx| and \lstinline|stat_awsim.xlsx|, which are the basis of Tab.~\ref{t:xp}.
The batch process covers a superset of the parameter grid described in Sect.~\ref{s:experiment}; it additionally includes $\Eps = 0.0125$ and $\Eps_v = 0.05$.

\end{document}